\documentclass[journal]{IEEEtran}
\usepackage{amsmath,amsfonts,amssymb}
\usepackage{algorithmic}
\usepackage{algorithm}
\usepackage{array}
\usepackage{textcomp}
\usepackage{url}
\usepackage{stfloats}  
\usepackage{subcaption} 

\usepackage{xcolor} 
\definecolor{darkgreen}{RGB}{24,224,24}
\usepackage[authormarkup=none]{changes} 
\definechangesauthor[color=blue]{me} 

\usepackage[colorlinks]{hyperref}
\usepackage{cite}
\usepackage{graphicx}
\graphicspath{{pics/}}
\newcommand{\figref}[1]{Fig.~\ref{#1}}
\usepackage{verbatim}  
\usepackage[utf8]{inputenc}

\begin{document}

\title{Spatial Semantic Communication: When Semantic Transmission Meets Index Modulation}

\author{
Xinghao Guo, Yin Xu,~\IEEEmembership{Senior Member,~IEEE,} Dazhi He,~\IEEEmembership{Senior Member,~IEEE,} \\ Hanjiang Hong,~\IEEEmembership{Member,~IEEE,} Zhiyong Chen,~\IEEEmembership{Senior Member,~IEEE,} Cixiao Zhang, \\Yiyan Wu,~\IEEEmembership{Life Fellow,~IEEE,} and Wenjun Zhang,~\IEEEmembership{Fellow,~IEEE} 

\thanks{Accepted by IEEE TCOM. \textit{(Corresponding author: Yin Xu.)}

X. Guo, Y. Xu, D. He, Z. Chen, C. Zhang and W. Zhang are with the Cooperative Medianet Innovation Center, Shanghai Jiao Tong University, Shanghai, 200240 China (e-mail: {guoxinghao, xuyin, hedazhi, zhiyongchen, cixiaozhang, zhangwenjun}@sjtu.edu.cn).

H. Hong is with the Department of Electronic and Electrical Engineering, University College London, Torrington Place, WC1E7JE, United Kingdom (e-mail: hanjiang.hong@ucl.ac.uk). 

Y. Wu is with the Department of Electrical and Computer Engineering, Western University, London, ON N6A 3K7 Canada (e-mail: yiyan.wu@ieee.org).

}
}

\markboth{Journal of \LaTeX\ Class Files,~Vol.~14, No.~8, August~20xx}%
{Shell \MakeLowercase{\textit{et al.}}: A Sample Article Using IEEEtran.cls for IEEE Journals}


\maketitle

\begin{abstract}
Current digital semantic communication  systems have primarily focused on maintaining compatibility with conventional constellation-based modulation. 
In contrast, index modulation (IM) represents a more spectrally and energy-efficient alternative by exploiting additional dimensions for information conveyance.
Recognizing this potential, this paper bridges the gap between IM and semantic communications by proposing a novel spatial semantic communication (SSC) system leveraging cutting-edge fluid antenna-IM (FA-IM) technology.
Compatible with existing joint source-channel coding (JSCC) architectures, the proposed SSC system employs the residual quantization (RQ) approach to discretize analog semantic features for subsequent digital IM transmission.
Notably, the proposed SSC system synergizes RQ and IM via a semantic-aware stream splitting scheme, which ensures that critical semantic information undergoes less severe channel fading, thereby further optimizing semantic transmission performance.
Simulation results validate that the proposed SSC system effectively integrates the high fidelity of RQ, the reliability of semantic-aware splitting, and the spatial efficiency of FA-IM, thereby providing a robust solution for future digital semantic transmission.
The open source code is available at: \url{https://github.com/gxh1106/SSC}.
\end{abstract}


\section{Introduction} \label{Sec-Introduction}
\IEEEPARstart{T}{he} deep synergy between sixth-generation (6G) communication and artificial intelligence (AI) has emerged as a defining trajectory of future wireless evolution, motivating novel intelligent communication paradigms \cite{ref-6G}.
In this context, semantic communication has garnered significant research attention for its superior coding gains and channel robustness \cite{ref-SemCom}. 
Unlike conventional systems, semantic communication aims to extract latent semantic features from the source, thereby ensuring the reliable transmission of information with fidelity at the semantic level.
Driven by the rapid advancements in deep learning (DL), semantic communications typically employ deep neural networks (DNNs) to implement an end-to-end joint source-channel coding (JSCC) framework.
This paradigm has been successfully extended to various data modalities, including text \cite{ref-TextSC}, image \cite{ref-SwinJSCC, ref-CDDM}, speech \cite{ref-SpeechSC}, video \cite{ref-VideoSC}, and even multimodal data \cite{ref-MMSC-2}.

Nevertheless, most existing semantic communication systems exhibit a critical incompatibility with modern digital communication infrastructure. 
The semantic features generated by DNN-based encoders are continuous-valued and designed for direct transmission in the analog domain. 
This approach poses significant hurdles in terms of hardware precision, practical deployment, and cost-effectiveness \cite{ref-DigitalSC}. 
Consequently, it is imperative to convert these encoded features into a finite set of discrete values. 
However, the direct application of traditional quantization methods risks the loss of critical semantic information. 

In this context, a substantial body of research has been dedicated to addressing the intricate challenge of the digital implementation of semantic communications.
One line of work focuses on mapping semantic features directly to discrete communication symbols, bypassing the bit conversion stage. 
For instance, the authors in \cite{ref-DeepJSCC-Q} proposed an end-to-end JSCC framework where the extracted semantic features are mapped onto a predefined constellation. 
Similarly, \cite{ref-L-DeepSC} projected the analog-valued outputs of the semantic encoder to a finite discrete constellation to enable lightweight deployment, whereas \cite{ref-SOM} introduced a multi-layer quantization design to mitigate quantization loss. 
In \cite{ref-JCM}, a joint coding-modulation framework was proposed that tackles the non-differentiability issue by leveraging the reparameterization trick. 
This scheme enables the network to learn the transition probability from source data to discrete constellation symbols.
More recently, a multi-order digital joint coding-modulation (MDJCM) scheme was developed in \cite{ref-R2-MDJCM} to directly integrate multi-order digital modulation into a joint coding framework.
A primary drawback of these direct-mapping approaches is the absence of an explicit bitstream, which renders them incompatible with existing bit-oriented digital communication protocols. 

To design bit-explicit semantic communication systems, several studies, such as \cite{ref-JSCC-Bit, ref-SparseSBC}, employed learnable quantizers to project features into a binary latent space. 
\cite{ref-R4-AMP} proposed an alternating multi-phase training strategy that incorporates mask-attack approximations to circumvent the non-differentiability, successfully enabling backward gradient propagation through discrete modulation and quantization bottlenecks.
\cite{ref-R5-ConcreteSC} proposed a fully learnable, multi-rate quantization framework which uses temperature-controlled concrete distributions, achieving multi-rate semantic transmission without retraining.
However, this approach is not only plagued by convergence instability, necessitating intricate training strategies, but also faces scalability challenges stemming from the exponential growth in parameter volume and computational complexity.

In contrast, an alternative research trajectory focusing on codebook-based digital JSCC has emerged as a more standard-compliant and computationally efficient paradigm. 
This approach draws inspiration from pioneering works in generative modeling, most notably the vector quantized-variational autoencoder (VQ-VAE) \cite{ref-VQ-VAE}.
Building upon this foundation, the systems developed in \cite{ref-DT-JSCC,ref-Masked-VQ-VAE} realize robust semantic communication, validating the remarkable resilience of the vector quantization (VQ) scheme against channel impairments. 
Specifically, the transmitter and receiver share a learnable codebook. 
At the transmitter, continuous semantic features are quantized into a sequence of codeword indices, which are then converted to bits for transmission. 
Subsequently, the receiver utilizes these indices to recover the feature vectors by retrieving the corresponding codewords from the shared codebook.
In a similar vein, the work in \cite{ref-VQ-DeepSC} proposed a multi-scale semantic extraction framework, complemented by a corresponding multi-scale VQ scheme.
Notably, the work in \cite{ref-sDAC} achieved efficient, bidirectional conversion between semantic features and bits by introducing a plug-and-play module. 
This module integrates VQ principles with a pair of learnable quantization adapters.
To optimize codebook assignment and channel adaptation, \cite{ref-R3-ESC-MVQ} developed an end-to-end framework leveraging multi-codebook VQ alongside iterative algorithms for adaptive modulation and power allocation.
Despite its effectiveness, this VQ-based paradigm still incurs substantial storage overhead and suffers from the codebook collapse problem, which hinders stable training \cite{ref-RQ-VAE}. 
To mitigate these issues, residual quantization (RQ), as introduced in residual-quantized VAE (RQ-VAE) \cite{ref-RQ-VAE}, adopts a multi-step quantization strategy to approximate the encoded features using a hierarchical stack of discrete codes. 
Building on this advancement, the work in \cite{ref-MOC-RVQ} leveraged the concept of RQ for image transmission, while \cite{ref-RVQGAN} employed it to achieve low-bitrate speech transmission.

While existing digital semantic communication frameworks have achieved commendable bit compatibility, the quest for higher spectral efficiency (SE) and energy efficiency (EE) remains a perpetual pursuit. 
This necessitates the continuous evolution of more spectrally efficient and sustainable physical layer technologies to meet the escalating demands of next-generation networks.
In this context, index modulation (IM) has emerged as a promising candidate, distinguishing itself by transcending the conventional modulation framework that relies exclusively on constellation symbols \cite{ref-IM-Survey}. 
IM unlocks new dimensions for information transmission by embedding data in the indices of physical layer resources, such as antennas \cite{ref-SM, ref-SM-NUC, ref-RIS-RGSM}, subcarriers \cite{ref-OFDM-IM}, and time slots.
More recently, the synergy between IM and fluid antenna (FA) technology \cite{ref-FAS} has been explored to further enhance SE.
Unlike conventional fixed-position antennas (FPAs), FA systems (FASs) offer unprecedented flexibility by enabling software-controlled modification of the antenna's physical properties within a predefined region, e.g., its position (also known as port) and shape, thus fully unlocking the spatial degrees of freedom \cite{ref-MA-Survey, ref-TNSE}. 
Capitalizing on this flexibility, the authors in \cite{ref-IM-FA} pioneered the FA-IM system, where the ports of the FA serve as the indexed entities to convey additional information, leading to significant improvements in both SE and bit error rate (BER) performance.
\cite{ref-FA-IM-NN} employed DNN to achieve fast classification of index patterns in FA-IM systems.
The work in \cite{ref-FA-PIM} proposed a position IM (PIM) system that enhances the overall performance of FAS by optimizing the port selection scheme.
\cite{ref-RIS-FA-IM} investigated the combination of FA-IM and reconfigurable intelligent surface (RIS)-assisted millimeter-wave (mmWave) communications.
\cite{ref-FA-IM} applied the FA-IM mechanism to MIMO systems, where multiple ports are activated simultaneously to map information to the index of port combination patterns, achieving enhanced SE.
The authors in \cite{ref-WCNC,ref-TCOM} proposed a novel FA grouping-based IM (FAG-IM) system to enhance robustness against spatial correlation, demonstrating superior BER performance.
Essentially, FASs provide a cost-effective solution to generate a massive number of ports. 
This abundance of spatial entities can be exploited by IM to convey additional information, thereby enhancing overall performance.

IM serves as a superior digital modulation paradigm by exploiting additional dimensions for information transmission. 
Although current digital semantic communications have established compatibility with conventional constellation-based modulation, integrating IM is essential to further push the boundaries of SE and EE. 
However, to the best of our knowledge, the intersection of semantic communication and IM remains an unexplored frontier. 
To fill this critical gap, this paper tackles the fundamental challenge of how to seamlessly fuse these two powerful paradigms. 
Specifically, we pioneer a novel communication paradigm named spatial semantic communication (SSC), exemplified by the state-of-the-art (SOTA) IM variant, FA-IM.
Designed as a generalized architecture compatible with existing analog semantic communication systems, the proposed SSC framework comprises three primary modules: a generic analog JSCC backbone, a residual quantizer with its corresponding dequantizer for digitization, and the FA-IM-based modulator and demodulator.
Crucially, building upon an in-depth analysis of both RQ and FA-IM, we propose a tailored semantic-aware stream splitting scheme to enhance data reconstruction quality.
Following the classical principle of unequal error protection (UEP), the proposed SSC system couples the hierarchical semantic significance of the source with the heterogeneous and asymmetric transmission reliability of the FA-IM streams.

The main contributions of this paper are summarized as follows:
\begin{itemize}
    \item \textbf{We propose an innovative SSC system}, which is the first framework to merge semantic communications and IM. 
    Compatible with conventional analog-domain JSCC backbones, the proposed system exploits the emerging FA-IM technology to convey semantic information via both FA port indices and constellation symbols.
    \item \textbf{We develop a learnable non-uniform quantizer and dequantizer} to bridge the gap between the analog JSCC architecture and digital FA-IM transmission.
    By incorporating a RQ approach, the system discretizes latent feature vectors into codeword indices via a shared codebook, thereby enhancing quantization fidelity.
    A dedicated loss function and a three-stage training strategy are proposed to ensure fast and stable end-to-end training.
    \item \textbf{We propose a novel semantic-aware stream splitting scheme}, whose design philosophy follows the classical principle of UEP, specifically tailored for the SSC system to further optimize transmission performance.
    This approach jointly considers the multi-granularity significance of semantic features produced by RQ and the unequal channel fading conditions experienced by the two data streams in the FA-IM framework. 
    By mapping critical semantic information to the more reliable data stream, our scheme establishes a synergistic coupling between RQ and FA-IM, thereby substantially enhancing reconstruction fidelity.
    \item \textbf{We perform extensive experiments} and results show that the proposed SSC system achieves superior performance compared to various benchmarks. 
    This validates the effectiveness of our joint design involving RQ, stream splitting, and FA-IM, demonstrating its robustness in enhancing reconstruction fidelity.   
\end{itemize}

The rest of this paper is organized as follows: Section \ref{Sec-SystemModel} presents the SSC system model and its internal signal transformation chain.
Section \ref{Sec-RQ} details the principles of RQ and the training strategy for the SSC system.
Section \ref{Sec-Split} introduces our core stream splitting scheme.
Section \ref{Sec-Simulation} shows the simulation and comparison results. 
Section \ref{Sec-Conclusion} concludes this paper.

\textit{Notations:} Scalar variables are denoted by italic letters, vectors are denoted by boldface small letters and matrices are denoted by boldface capital letters. $(\cdot)^*$ denotes the conjugate operation of a complex scalar variable. $\mathrm{det}(\cdot)$ stands for the determinant while $(\cdot)^T$, $(\cdot)^{-1}$ and $(\cdot)^H$ denote transposition, inverse and Hermitian transposition of a matrix, respectively. $| \cdot |$ and ${\| \cdot \|}_2$ denote the absolute and the $\ell_2$ norm operations, respectively. $\binom{\cdot}{\cdot}$ and $\lfloor \cdot \rfloor$ denote the binomial coefficient and the floor operation, respectively. $\otimes$ denotes the Kronecker product, and $\mathrm{vec}(\cdot)$ denotes the vectorization operator. $\mathrm{diag}(\cdot)$ denotes a diagonal matrix whose diagonal entries are the inputs. $\mathbb{E}[\cdot]$ returns the expected value of the input random quantity. The real and imaginary parts of a complex variable $X$ are denoted by $\Re\{X\}$ and $\Im\{X\}$.

\begin{figure*}[t]
\centerline{\includegraphics[width=0.75\textwidth]{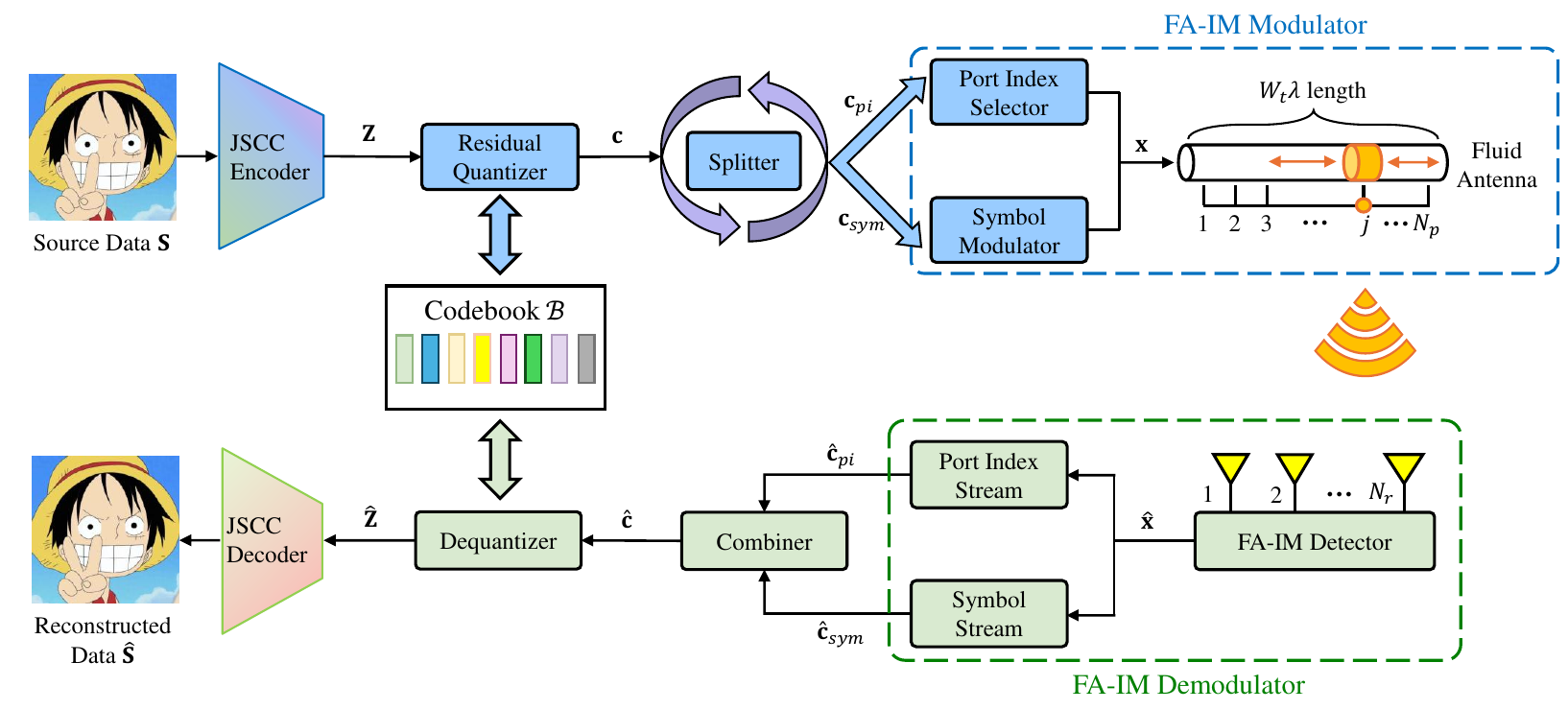}}
\captionsetup{justification=justified, singlelinecheck=false}
\caption{Block diagram of the proposed SSC system.}
\label{fig-SSC_system}
\end{figure*}

\section{Proposed System Model} \label{Sec-SystemModel}
The architecture of the proposed SSC system is depicted in \figref{fig-SSC_system}.
The system integrates a JSCC backbone with a RQ module, alongside an FA-IM modulator and a corresponding demodulator.
The transmitter is equipped with a single FA, whose position can be switched instantaneously to one of $N_p$ uniformly distributed ports along a one-dimensional (1D) linear space of length $W_t \lambda$, where $\lambda$ is the wavelength of radiation and $W_t$ denotes the length of the FA normalized by $\lambda$. 
Consequently, the spacing between adjacent ports is given by $\Delta_t= \frac{W_t \lambda}{N_p - 1}$.
The receiver employs a conventional uniform linear array (ULA) composed of $N_r$ FPAs with an inter-element spacing of $\Delta_r= \lambda/2$.

\subsection{Channel Model} \label{Sec-SystemModel-ChannelModel}
In a finite-scattering environment, such as that of millimeter-wave (mmWave) communication systems, the channel can be effectively characterized by the planar-wave geometric model \cite{ref-MA}.
Neglecting the path loss, since its effects are assumed to be accounted for by the received signal-to-noise ratio (SNR), the channel matrix $\mathbf{H} \in \mathbb{C}^{N_r \times N_p}$ can be expressed as
\begin{equation}
\label{eq-channel}
\mathbf{H}=\sqrt{\frac{N_p N_r}{L_p}} \sum_{l=1}^{L_p} \alpha_l \mathbf{a}_r(\vartheta_l) \mathbf{a}_t^H(\varphi_l),
\end{equation}%
where $L_p$ denotes the number of propagation paths, $\alpha_l \sim \mathcal{CN}(0, 1)$ is the response coefficient of the $l$-th path, $\varphi_l \in [-\frac{\pi}{2}, \frac{\pi}{2}]$ and $\vartheta_l \in [-\frac{\pi}{2}, \frac{\pi}{2}]$ are the angle of departure (AoD) and angle of arrival (AoA) for the $l$-th path, respectively, and $\mathbf{a}_t(\varphi_l) \in \mathbb{C}^{N_p \times 1}$ and $\mathbf{a}_r(\vartheta_l) \in \mathbb{C}^{N_r \times 1}$ denote the corresponding transmit and receive steering vectors, respectively, defined as follows:
\begin{equation}
\begin{aligned}
\label{eq-steeringvector}
\mathbf{a}_t(\varphi_l) &= \sqrt{\frac{1}{N_p}}\left[1, e^{j \frac{2 \pi}{\lambda} \Delta_t \sin (\varphi_l)}, \ldots, e^{j \frac{2 \pi}{\lambda} \Delta_t (N_p-1) \sin (\varphi_l)}\right]^T, \\
\mathbf{a}_r(\vartheta_l) &= \sqrt{\frac{1}{N_r}}\left[1, e^{j \frac{2 \pi}{\lambda} \Delta_r \sin (\vartheta_l)}, \ldots, e^{j \frac{2 \pi}{\lambda} \Delta_r (N_r-1) \sin (\vartheta_l)}\right]^T.
\end{aligned}
\end{equation}%

To implement IM, the transmitter pre-selects $N_s$ out of $N_p$ ports, with $N_s$ being a power of two.
The set of selected port indices is denoted by $\mathcal{I}$ ($|\mathcal{I}| = N_s, \mathcal{I} \subseteq \{1, ..., N_p\}$), which in turn defines the sub-channel matrix $\mathbf{H}_{\mathcal{I}}$ composed of the corresponding columns from $\mathbf{H}$.
Assuming perfect channel state information (CSI) at the transmitter and adopting a capacity maximization goal similar to \cite{ref-MIMO-FAS}, the optimal sub-channel matrix $\bar{\mathbf{H}}$ is found by
\begin{equation}
\begin{aligned}
\label{eq-optimalchannel}
    \bar{\mathbf{H}} = \arg \underset{\mathbf{H}_{\mathcal{I}}}{\max} \ \mathcal{C}(\mathbf{H}_{\mathcal{I}}),
\end{aligned}
\end{equation}%
where $\mathcal{C}(\mathbf{H}_{\mathcal{I}})$ is the capacity, computed as 
\begin{equation}
\label{eq-channelcapacity}
    \mathcal{C}(\mathbf{H}_{\mathcal{I}}) = \log_2 \mathrm{det} \left( \mathbf{I}_{N_r} + \frac{1}{N_s} \mathbf{H}_{\mathcal{I}} \mathbf{H}_{\mathcal{I}}^H \right),
\end{equation}%
where $\mathbf{I}_{N_r}$ is the identity matrix of size $N_{r}$.
$\bar{\mathbf{H}}$ can be obtained either through an exhaustive search over all $\binom{N_p}{N_s}$ possible combinations, or by employing a conventional low-complexity greedy algorithm.

\subsection{Signal Model} \label{Sec-SystemModel-SignalModel}
This subsection presents the end-to-end signal processing flow within the proposed SSC system. 

\subsubsection{Transmitter}
On the transmitter side, the source data $\mathbf{S}$ is encoded by a JSCC encoder, $Enc(\cdot)$, into a low-dimensional latent representation $\mathbf{Z} = Enc(\mathbf{S}) \in \mathbb{R}^{L_z \times C_z}$. Here, $L_z$ denotes the length of the latent sequence, and $C_z$ represents the dimensionality of each latent vector.
The semantic feature $\mathbf{Z}$ consists of continuous scalar values. 
For digital transmission, a residual quantizer $RQ(\cdot)$ operates element-wise to map each value in $\mathbf{Z}$ to a sequence of codeword indices, denoted as $\mathbf{c}$. 
This process leverages a learnable codebook $\mathcal{B}$, shared between the transmitter and receiver, and is expressed as $\mathbf{c} = RQ(\mathbf{Z}; \mathcal{B})$.
A comprehensive discussion about the design of $RQ(\cdot)$ and the associated training strategies is provided in Section \ref{Sec-RQ}.
Next, a stream splitter divides the index sequence $\mathbf{c}$ into a port index stream, denoted as $\mathbf{c}_{pi}$, and a constellation symbol stream, denoted as $\mathbf{c}_{sym}$, for subsequent FA-IM mapping. 
Herein lies a key contribution of this work, which is detailed in Section \ref{Sec-Split}: a dedicated splitting scheme is proposed to organically couple the characteristics of RQ with the FA-IM mechanism, thereby further boosting the SSC system performance.

Within the FA-IM modulator, two processes occur in parallel during each transmission slot. 
The port index selector first converts the stream $\mathbf{c}_{pi}$ into a bit sequence, which is then segmented into blocks, with each block containing $m_1 = \log_2 N_s$ bits.
Sequentially, each block is utilized to determine a specific port index $j$ within the pre-selected set $\mathcal{I}$ of size $N_s$, where $\mathcal{I}$ is obtained via \eqref{eq-optimalchannel}. 
Simultaneously, the symbol modulator performs a similar procedure. 
It converts the stream $\mathbf{c}_{sym}$ into a bit sequence, segments it, and maps each block of $m_2 = \log_2 M$ bits to a symbol $s$ using an $M$-ary constellation alphabet $\mathcal{S}$.
As a result, the $j$-th port is activated and transmits the symbol $s$.
Therefore, the transmitted signal vector $\mathbf{x} \in \mathbb{C}^{N_p}$ at the SSC transmitter can be expressed as 
\begin{equation}
\label{eq-transmittedsignal}
\mathbf{x} = s \mathbf{v}_j, 
\end{equation}%
where $\mathbf{v}_j$ is a standard basis vector of dimension $N_p$ with a one at the $j$-th position and zeros elsewhere.
The corresponding SE, in terms of bits per channel use (bpcu), is given by
\begin{equation}
\label{eq-SE_SSC}
\mathrm{SE_{\text{SSC}}}= m_1 + m_2 = \log_2 N_s + \log_2 M \ [\mathrm{bpcu}].
\end{equation}%

\subsubsection{Receiver}
At the receiver, the received signal $\mathbf{y} \in \mathbb{C}^{N_r}$ can be written as
\begin{equation}
\label{eq-receivedsignal}
\mathbf{y}= \bar{\mathbf{H}} \mathbf{x} + \mathbf{n},
\end{equation}%
where $\mathbf{n} \in \mathbb{C}^{N_r} \sim \mathcal{CN}(0, N_0 \mathbf{I}_{N_r})$ denotes the additive white Gaussian noise (AWGN) vector.
Under the assumption of perfect CSI at the receiver, the optimal maximum likelihood (ML) detector performs an exhaustive search over the $N_s$ candidate port indices and $M$ possible symbols, which can be expressed as 
\begin{equation}
\begin{aligned}
\label{eq-MLdetector}
    (\hat{j},\hat{s})&= \arg \underset{j \in \mathcal{I}, s  \in \mathcal{S}}{\min} \left \| \mathbf{y} - \bar{\mathbf{H}} \mathbf{x} \right \|_2^2\\
    &=\arg \underset{j \in \mathcal{I}, s  \in \mathcal{S}}{\min} \left \| \mathbf{y} - s \bar{\mathbf{h}}_j \right \|_2^2,
\end{aligned}
\end{equation}%
where $\bar{\mathbf{h}}_j$ represents the $j$-th column of $\bar{\mathbf{H}}$.

Upon obtaining the detected pair $(\hat{j},\hat{s})$, the FA-IM demodulator recovers the corresponding index and symbol substreams.
These substreams are then fed into a stream combiner, which performs the inverse operation of the splitting scheme used at the transmitter to yield the estimate $\hat{\mathbf{c}}$, potentially corrupted by detection errors.
Then, the estimated sequence $\hat{\mathbf{c}}$ is fed into a dequantizer $DQ(\cdot)$, which is the counterpart to the residual quantizer $RQ(\cdot)$ at the transmitter. 
Utilizing the shared codebook $\mathcal{B}$, the dequantizer reproduces the semantic feature $\hat{\mathbf{Z}}$, which is expressed as $\hat{\mathbf{Z}} = DQ(\hat{\mathbf{c}}; \mathcal{B})$.
Finally, based on $\hat{\mathbf{Z}}$, the JSCC decoder $Dec(\cdot)$ generates a high-quality reconstruction of the source data, $\hat{\mathbf{S}}$.

In summary, the complete signal chain within the SSC system is as follows:
\begin{equation}
\label{eq-SSCflow}
\begin{aligned}
    \mathbf{S} \xrightarrow{Enc(\cdot)} \mathbf{Z} \xrightarrow{RQ(\cdot; \mathcal{B})} \mathbf{c} \xrightarrow{\text{Splitter}} (\mathbf{c}_{pi}, \mathbf{c}_{sym})  \xrightarrow{\text{FA-IM Mod}} & \mathbf{x} \\
    & \downarrow \bar{\mathbf{H}} \\
    \hat{\mathbf{S}} \xleftarrow{Dec(\cdot)} \hat{\mathbf{Z}} \xleftarrow{DQ(\cdot; \mathcal{B})} \hat{\mathbf{c}} \xleftarrow{\text{Combiner}} (\hat{\mathbf{c}}_{pi}, \hat{\mathbf{c}}_{sym}) \xleftarrow{\text{FA-IM Demod}} & \mathbf{y}
\end{aligned}
\end{equation}%

\begin{figure*}[t]
    \centering
    \begin{subfigure}[b]{0.7\textwidth} 
        \centering
        \includegraphics[width=1\linewidth]{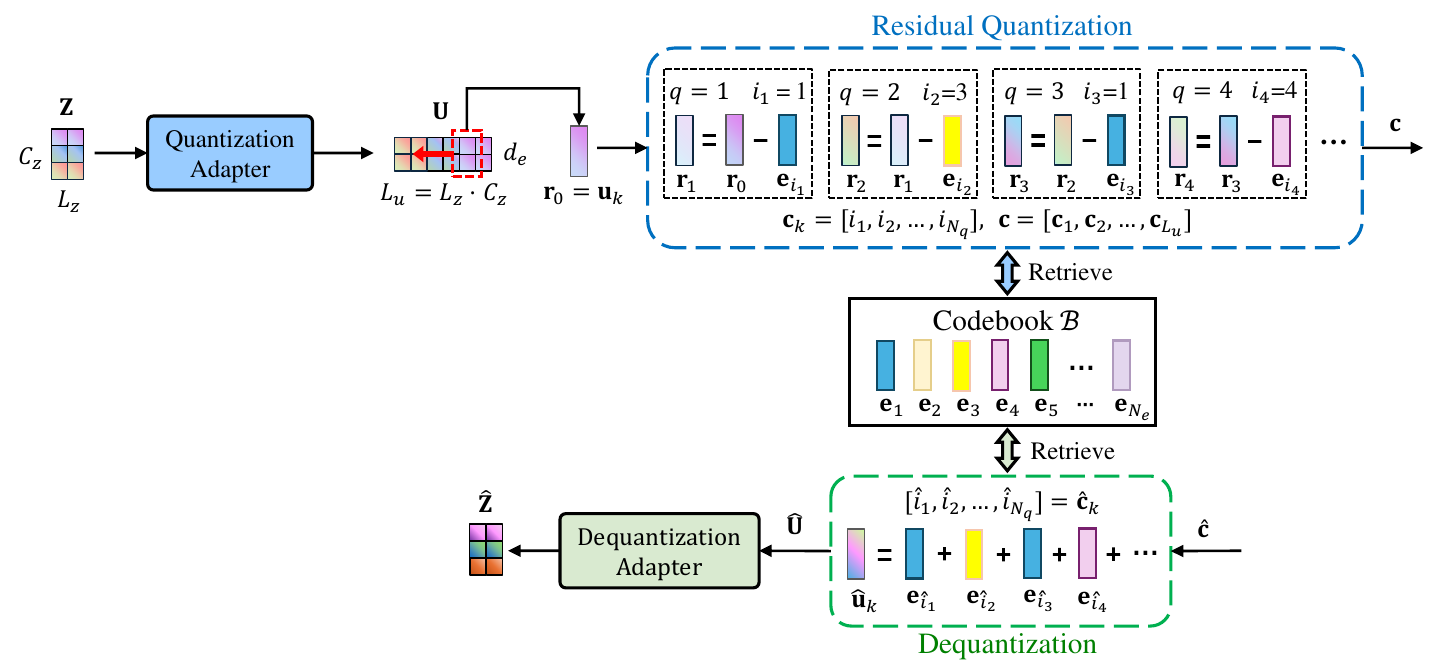}
        \subcaption{Residual Quantizer and Dequantizer}
        \label{fig-RQmodule-overview}
    \end{subfigure}
    \hspace{-1cm} 
    \begin{minipage}[b]{0.32\textwidth}
        \centering
        \begin{subfigure}{\linewidth} 
            \centering
            \includegraphics[width=\linewidth]{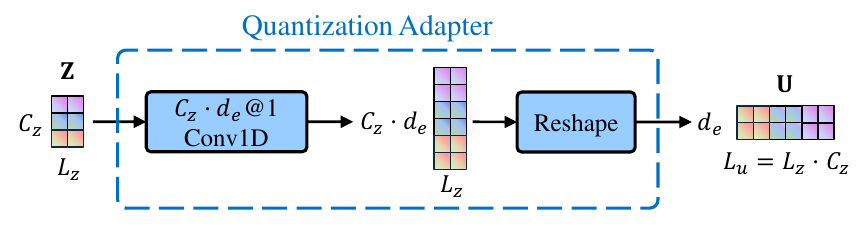}
            \subcaption{Quantization Adapter}
            \label{fig-RQmodule-quantadapt}
        \end{subfigure}
        
        \vspace{10pt} 
        
        \begin{subfigure}{\linewidth}
            \centering
            \includegraphics[width=0.8\linewidth]{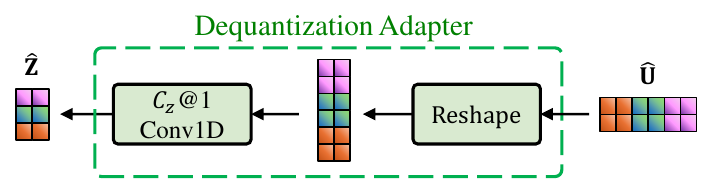}
            \subcaption{Dequantization Adapter}
            \label{fig-RQmodule-dequantadapt}
        \end{subfigure}
    \end{minipage}

    \caption{Block diagrams of the proposed residual quantizer and dequantizer in the SSC system. (a) Overview of the residual quantizer and dequantizer. (b) Architecture of the quantization adapter. (c) Architecture of the dequantization adapter.}
    \label{fig-RQmodule}
\end{figure*}

\subsection[Limitations]{Limitations} \label{Sec-SystemModel-limit}
Although the proposed SSC system is designed and evaluated with FA-IM, we emphasize that the core contributions of this work are inherently generalizable. 
Since these digitization and semantic-aware stream splitting designs operate at the index and bit mapping levels, they are mathematically decoupled from the physical layer antenna architecture. 
Consequently, they can be seamlessly applied to other classical or emerging IM paradigms, such as spatial modulation (SM) \cite{ref-SM} and subcarrier index modulation \cite{ref-OFDM-IM}. 
In this paper, FA-IM is employed primarily as a SOTA representative paradigm to demonstrate the efficacy and robustness of our joint design.

Furthermore, we address a practical limitation concerning the physical switching latency of fluid antennas.
While mechanical- or liquid-based FA implementations may suffer from mechanical inertia, modern pixel-based reconfigurable fluid antennas utilize solid-state electronic switches (e.g., PIN diodes or RF-MEMS) to dynamically activate target ports without physical movement, thereby successfully pushing the port-switching latency down to the microsecond level \cite{ref-pixel-FA}. 
It is also worth noting that switching latency is not a unique drawback of FA-IM, but a fundamental bottleneck shared by the entire IM family due to the frequent toggling of active antennas or RF-chains. 
Fortunately, several advanced transmission designs have been established in the literature to mitigate this issue. 
A prominent example is offset SM \cite{ref-OSM}, which introduces structured offsets to drastically reduce the RF chain switching frequency, or even eliminate the switching overhead entirely. 
Such latency-reduction schemes can be naturally integrated with our proposed SSC framework in future extensions to further enhance the real-time processing capability of the system.

\section{Residual Quantization and Training Methods} \label{Sec-RQ}
This section elaborates on the architecture and operational mechanism of the proposed quantizer and the corresponding dequantizer in the SSC system.
Following this, the end-to-end training strategies for the SSC system are detailed.

\subsection{Residual Quantization} \label{Sec-RQ-RQ}
\figref{fig-RQmodule-overview} illustrates the overall architecture of the proposed residual quantizer $RQ(\cdot)$ and its corresponding dequantizer $DQ(\cdot)$. 
$RQ(\cdot)$ is responsible for mapping the $L_z C_z$ continuous values within the latent feature tensor $\mathbf{Z}$ to sequences of codeword indices from the shared codebook $\mathcal{B}$.
Conversely, $DQ(\cdot)$ reconstructs the latent feature by retrieving the corresponding codewords from $\mathcal{B}$ based on the received index sequence.

To this end, the input tensor $\mathbf{Z}$ is first projected by a quantization adapter, inspired by \cite{ref-sDAC}, into a new representation $\mathbf{U} = [\mathbf{u}_1, \mathbf{u}_2, \ldots, \mathbf{u}_{L_u}]^T \in \mathbb{R}^{L_u \times d_e}$. 
Here, $L_u = L_z C_z$ corresponds to the total number of values to be quantized from $\mathbf{Z}$, and $d_e$ denotes the dimensionality of each vector $\mathbf{u}_k, k=1, 2, \ldots, L_u$. 
The dimension $d_e$ is set to match the dimensionality of the codewords in the codebook $\mathcal{B}$, facilitating the subsequent quantization process.
As illustrated in \figref{fig-RQmodule-quantadapt}, the quantization adapter is composed of a 1D convolutional layer followed by a reshape operation. 
The convolutional layer, which has a kernel size of 1, expands the input tensor channels from $C_z$ to $C_z  d_e$. The resulting tensor is then reshaped to the target dimensions of $L_u \times d_e$.

Next, departing from the VQ approach used in \cite{ref-sDAC}, our SSC system employs the RQ approach to discretize $\mathbf{U}$ into the index sequence $\mathbf{c}$.
Specifically, the shared codebook $\mathcal{B}$, comprising $N_e$ codewords, is defined as $\mathcal{B} = \{\mathbf{e}_i \in \mathbb{R}^{d_e} \mid i = 1, 2, \ldots, N_e \}$, where $\mathbf{e}_i$ denotes the $i$-th codeword in the codebook. 
With $N_q$ denoting the number of quantization steps, each input vector $\mathbf{u}_k$ is ultimately mapped to an index sequence $\mathbf{c}_k$, which belongs to the space $\{1, 2, \ldots, N_e\}^{N_q}$.
Starting with an initial residual $\mathbf{r}_0 = \mathbf{u}_k$, the RQ process iteratively computes the codeword indices by finding the codeword in $\mathcal{B}$ that has the minimum Euclidean distance to the current residual vector. 
Thus, the operation at the $q$-th quantization step is formulated as
\begin{equation}
\begin{aligned}
\label{eq-RQ}
    &i_q = \arg \underset{i \in \{1, 2, \ldots, N_e \}}{\min} \left \| \mathbf{r}_{q-1} - \mathbf{e}_i \right \|_2^2, \\
    &\mathbf{r}_q = \mathbf{r}_{q-1} - \mathbf{e}_{i_q},
\end{aligned}
\end{equation}%
where $i_q$ is the obtained codeword index and $\mathbf{r}_q$ is the updated residual at step $q$, for $q=1, 2, \ldots, N_q$.
After $N_q$ quantization steps, each vector $\mathbf{u}_k$ is discretized into the index sequence $\mathbf{c}_k = [i_1, i_2, \ldots, i_{N_q}]$. 
Here, the full notation for each index is $i_{k,1}, i_{k,2}, \dots, i_{k,N_q}$, we omit the subscript $k$ for simplicity.
Consequently, the final output for the entire input $\mathbf{U}$ is the concatenation of the individual sequences, yielding $\mathbf{c} = [\mathbf{c}_1, \mathbf{c}_2, \ldots, \mathbf{c}_{L_u}] \in \{1, 2, \ldots, N_e\}^{L_u N_q}$.
To facilitate subsequent bit-based digital transmission, $N_e$ is recommended to be set to a power of 2.

The RQ approximates each input vector $\mathbf{u}_k$ in a coarse-to-fine fashion. 
With each additional quantization step, the quantization error is progressively reduced. 
In other words, by increasing the number of quantization steps $N_q$, the cumulative sum of the selected codewords, $\sum_{q=1}^{N_q} \mathbf{e}_{i_q}$, forms an increasingly precise approximation of the original vector. 
Notably, for the case of $N_q=1$, the RQ scheme degenerates into standard VQ. 
The critical limitation of VQ, therefore, is that improving precision relies solely on enlarging the codebook size $N_e$. 
However, this requires $N_e$ to grow exponentially, which leads to the codebook collapse issue and renders the training process extremely unstable.
In summary, for a given codebook size, RQ achieves superior approximation accuracy over VQ in an efficient manner.

The index sequence $\mathbf{c}$ output by $RQ(\cdot)$ undergoes wireless transmission via FA-IM, resulting in the received sequence $\hat{\mathbf{c}}$, which may contain errors.
The dequantizer $DQ(\cdot)$ partitions $\hat{\mathbf{c}}$ into $L_u$ individual index sequences $\hat{\mathbf{c}}_k, k=1, 2, \ldots, L_u$, each of length $N_q$. 
For each sequence $\hat{\mathbf{c}}_k= [\hat{i}_1, \hat{i}_2, \ldots, \hat{i}_{N_q}]$, the corresponding vector $\hat{\mathbf{u}}_k$ is reconstructed by summing the $N_q$ indicated codewords retrieved from the codebook $\mathcal{B}$, which is expressed as
\begin{equation}
\label{eq-recon_codewords}
\hat{\mathbf{u}}_k = \sum_{q=1}^{N_q} \mathbf{e}_{\hat{i}_q}. 
\end{equation}%
The vectors $\hat{\mathbf{u}}_k$ are assembled to form $\hat{\mathbf{U}}$ and fed into the dequantization adapter, whose architecture is shown in \figref{fig-RQmodule-dequantadapt}. 
The adapter performs the inverse operation of its counterpart at the transmitter. 
Specifically, $\hat{\mathbf{U}}$ is first reshaped and then passed through a 1D convolutional layer with kernel size 1. 
This layer compresses the feature channels from $C_z d_e$ back down to $C_z$, yielding the final reconstructed latent feature $\hat{\mathbf{Z}}$.

\subsection{End-to-End Training Framework} \label{Sec-RQ-Training}
\subsubsection{Binary Symmetric Channel}
Simulating the full FA-IM transmission link, as described in \eqref{eq-SSCflow}, during the training phase would introduce prohibitive computational complexity and significantly prolong the training time. 
Crucially, we note that the adversarial effect of the channel on the transmitted semantic information during training is fundamentally manifested as bit errors. 
Therefore, to facilitate an efficient end-to-end training process, we adopt the widely-used binary symmetric channel (BSC) model as a substitute for the actual FA-IM transmission \cite{ref-JSCC-Bit}. 
The BSC provides a simple yet effective way to simulate channel-induced errors, thereby simplifying the training pipeline.
Specifically, during forward propagation, the index sequence $\mathbf{c}$ is first converted from decimal to binary form to obtain a bit sequence $\mathbf{b}$ of length $L_u N_q \log_2 N_e$. 
Subsequently, each bit in $\mathbf{b}$ undergoes an independent bit-flip operation with a given probability $p$, yielding the erroneous bit sequence $\hat{\mathbf{b}}$, which is formulated as:
\begin{equation}
\label{eq-BSC}
\hat{\mathbf{b}}=
\begin{cases}
\mathbf{1}-\mathbf{b}, & \text{w.p. } p \\
\mathbf{b}, & \text{w.p. } 1-p
\end{cases}.
\end{equation}%
The receiver module then converts $\hat{\mathbf{b}}$ from binary back to decimal to reconstruct the index sequence $\hat{\mathbf{c}}$.

It is worth noting that while real-world finite-scattering mmWave environments experience correlated fading, their combined physical impairments ultimately manifest as bit-flip errors on the digitized bitstream at the decoder input. 
By employing the BSC, we abstract the complex physical communication process into equivalent transition probabilities. 
To ensure that the trained neural network generalizes robustly to these varying, correlated physical channel states during evaluation, the bit-flip probability $p$ is not kept constant. 
Instead, for each forward training pass, $p$ is randomly sampled from a wide range of predefined values calculated across uniformly spaced SNR levels. 
This dynamic training strategy exposes the network to a diverse spectrum of error patterns, thereby forcing the semantic decoder to learn a highly generalized denoising mapping that robustly accommodates the actual bit error distributions produced by the physical fading channels.

\subsubsection{Training Objectives}
The training objective of the proposed SSC system is to optimize the end-to-end semantic communication performance. 
Our system is designed to be compatible with existing JSCC backbones, allowing for task-specific loss functions depending on the semantic task at hand. 
Without loss of generality, this paper considers the image reconstruction task as an illustrative example. 
Accordingly, the loss function is defined as the mean squared error (MSE) between the original and reconstructed images:
\begin{equation}
\label{eq-loss_recon}
\mathcal{L}_{\text{recon}} = ||\mathbf{S} - \hat{\mathbf{S}}||_2^2. 
\end{equation}%

In addition to the primary semantic task loss, the overall objective function incorporates a commitment loss, $\mathcal{L}_{\text{commit}}$, to mitigate RQ errors, which is defined as
\begin{equation}
\label{eq-loss_commit}
\mathcal{L}_{\text{commit}} = \sum_{k=1}^{L_u} \sum_{q=1}^{N_q} \left \| \mathbf{u}_k - \text{sg}\left [\sum_{t=1}^{q}\mathbf{e}_{i_t} \right ]\right \|_2^2, 
\end{equation}%
where sg[$\cdot$] denotes the stop-gradient operation, and the term $\sum_{t=1}^{q}\mathbf{e}_{i_t}$ is the reconstruction of $\mathbf{u}_k$ using the codewords selected up to step $q$. 
Notably, the commitment loss is designed to accumulate the quantization error at each intermediate step $q$, rather than merely calculating the final error after all $N_q$ steps, i.e., $\sum_{k=1}^{L_u} \| \mathbf{u}_k - \text{sg} [\sum_{q=1}^{N_q}\mathbf{e}_{i_q}  ] \|_2^2$. 
This design choice explicitly encourages the RQ model to reduce the quantization error sequentially as the quantization stage $q$ progresses.

To ensure a smooth and stable training process for the codebook $\mathcal{B}$ and to prevent the issue of codebook collapse, we update $\mathcal{B}$ using an exponential moving average (EMA) scheme instead of conventional gradient descent.
The update rule is given by
\begin{equation}
\label{eq-codebook-ema}
\mathbf{e}_i \leftarrow \gamma \mathbf{e}_i + (1 - \gamma) \bar{\mathbf{u}}_i, \ i = 1, 2, \ldots, N_e
\end{equation}%
where $\gamma \in [0, 1)$ is the decay factor, and $\bar{\mathbf{u}}_i$ represents the mean of all input vectors $\{\mathbf{u}_k\}$ from the current batch that are mapped to the codeword $\mathbf{e}_i$. 

In summary, the final loss function for the proposed SSC system is formulated as
\begin{equation}
\label{eq-loss_total}
\mathcal{L}_{\text{total}} = \mathcal{L}_{\text{recon}} + \beta \mathcal{L}_{\text{commit}}, 
\end{equation}%
where the hyperparameter $\beta$ controls the weight of the commitment loss.

Furthermore, to address the non-differentiability of the nearest neighbor operation in RQ, as shown in \eqref{eq-RQ}, the straight-through estimator (STE) technique is employed for gradient backpropagation. 
Specifically, we define a differentiable proxy vector $\mathbf{u}_{k, \text{STE}}$ for each continuous latent vector $\mathbf{u}_k \in \mathbb{R}^{d_e}$ as
\begin{equation}
\label{eq-STE}
\mathbf{u}_{k, \text{STE}} = \mathbf{u}_k + \text{sg}[\hat{\mathbf{u}}_k - \mathbf{u}_k].
\end{equation}%
During forward propagation, the identity property of the stop-gradient operator yields $\mathbf{u}_{k, \text{STE}} = \mathbf{u}_k + (\hat{\mathbf{u}}_k - \mathbf{u}_k) = \hat{\mathbf{u}}_k$. 
This ensures that the downstream JSCC decoder $Dec(\cdot)$ operates on the exact discrete reconstructed features $\hat{\mathbf{u}}_k$.
During backward propagation, since the derivative of the stop-gradient operator is zero, we can evaluate the gradient of the loss $\mathcal{L}_{\text{recon}}$ with respect to the input latent feature $\mathbf{u}_k$ using the chain rule:
\begin{equation}
\label{eq-chainrule}
\frac{\partial \mathcal{L}_{\text{recon}}}{\partial \mathbf{u}_k} = \frac{\partial \mathcal{L}_{\text{recon}}}{\partial \mathbf{u}_{k, \text{STE}}} \cdot \frac{\partial \mathbf{u}_{k, \text{STE}}}{\partial \mathbf{u}_k}.
\end{equation}%
For the first term, it is calculated as:
\begin{equation}
\begin{aligned}
\label{eq-chainrule-first}
\frac{\partial \mathcal{L}_{\text{recon}}}{\partial \mathbf{u}_{k, \text{STE}}} &= \frac{\partial \|Dec(\mathbf{u}_{k, \text{STE}}) - \mathbf{S}\|_2^2}{\partial \mathbf{u}_{k, \text{STE}}} \\
&= 2(Dec(\mathbf{u}_{k, \text{STE}}) - \mathbf{S})^T \cdot \frac{\partial Dec(\mathbf{u}_{k, \text{STE}})}{\partial \mathbf{u}_{k, \text{STE}}}.
\end{aligned}
\end{equation}
Since the derivative of the stop-gradient operator is zero, the second term is denoted as
\begin{equation}
\label{eq-chainrule-second}
\frac{\partial \mathbf{u}_{k, \text{STE}}}{\partial \mathbf{u}_k} = \frac{\partial \mathbf{u}_k}{\partial \mathbf{u}_k} + \frac{\partial \text{sg}[\hat{\mathbf{u}}_k - \mathbf{u}_k]}{\partial \mathbf{u}_k} = \mathbf{I} + \mathbf{0} = \mathbf{I},
\end{equation}
where $\mathbf{I}$ represents the identity matrix. 
Therefore, the gradient in \eqref{eq-chainrule} can be written as:
\begin{equation}
\begin{aligned}
\label{eq-gradient}
\frac{\partial \mathcal{L}_{\text{recon}}}{\partial \mathbf{u}_k} &= 2(Dec(\mathbf{u}_{k, \text{STE}}) - \mathbf{S})^T \cdot \frac{\partial Dec(\mathbf{u}_{k, \text{STE}})}{\partial \mathbf{u}_{k, \text{STE}}} \cdot \mathbf{I}\\
&= 2(Dec(\hat{\mathbf{u}}_k) - \mathbf{S})^T \cdot \frac{\partial Dec(\hat{\mathbf{u}}_k)}{\partial \hat{\mathbf{u}}_k}.
\end{aligned}
\end{equation}%
This mathematical formulation confirms that the gradient of the reconstruction loss can successfully bypass the non-differentiable $N_q$-step quantization and flow directly back to the continuous feature $\mathbf{u}_k$, enabling seamless end-to-end optimization.

\begin{figure}[t]
\centerline{\includegraphics[width=0.7\columnwidth]{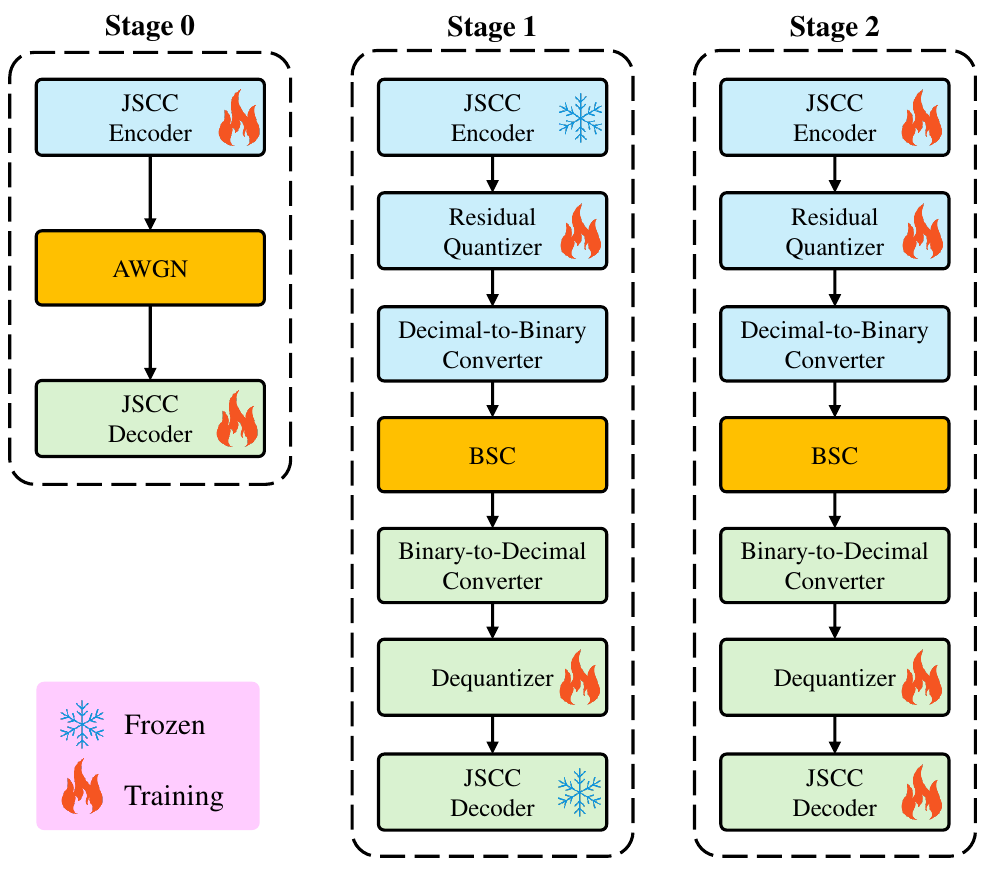}}
\captionsetup{justification=justified, singlelinecheck=false}
\caption{The multi-stage training processes for SSC.}
\label{fig-SSC_training}
\end{figure}
\subsubsection{Training Strategies}
We propose a three-stage training strategy to ensure training stability and enhance the robustness of the SSC system, as illustrated in \figref{fig-SSC_training}. 
In Stage 0, we train the backbone of the SSC system, which consists of the encoder $Enc(\cdot)$ and the decoder $Dec(\cdot)$. 
This stage aims to strengthen the semantic compression and reconstruction capabilities of the backbone in the analog domain. 
To fortify the backbone against real-world channel noise and potential quantization errors, we employ a noise-injection method adopted in \cite{ref-SwinJSCC}: during each forward pass, an SNR value is randomly sampled from a predefined set of candidates, and AWGN of the corresponding level is added to the analog semantic features output by the encoder $Enc(\cdot)$. 
Furthermore, enabled by the inherent compatibility of our SSC system with standard JSCC frameworks, the pre-training in Stage 0 is entirely optional.
One can directly initialize the backbone with an off-the-shelf, pre-trained JSCC model, thereby significantly accelerating the training pipeline.

In the subsequent Stage 1, the residual quantizer $RQ(\cdot)$ and dequantizer $DQ(\cdot)$ are integrated into the network for training, with the codebook $\mathcal{B}$ being updated via EMA. 
Differing from the complete SSC system which operates over the FA-IM channel, the output of $RQ(\cdot)$ at this stage is passed through the BSC, whose bit error probability $p$ is randomly sampled for each forward pass, mirroring the manner of Stage 0. 
Concurrently, the pre-trained encoder $Enc(\cdot)$ and decoder $Dec(\cdot)$ are frozen. 
This is crucial to prevent them from being destabilized by the large, erratic gradients from the newly initialized RQ modules, thereby avoiding catastrophic forgetting of the learned analog representation. 
The objective of this stage is thus the initial alignment of the RQ modules with the analog semantic feature space.

In the final Stage 2, the encoder $Enc(\cdot)$ and decoder $Dec(\cdot)$ are unfrozen to allow for a global fine-tuning of the entire SSC network. 
The primary objective is to achieve a final, deep alignment among the analog JSCC modules, the RQ modules, and the discrete codebook. 
Moreover, this joint optimization allows the SSC system to leverage the powerful reconstruction capabilities of the decoder $Dec(\cdot)$ to further enhance robustness against both channel noise and quantization errors, boosting the overall resilience and fidelity.

\begin{figure}[t]
    \centering
    \begin{subfigure}{0.48\columnwidth}
        \centering
        \includegraphics[width=1\linewidth]{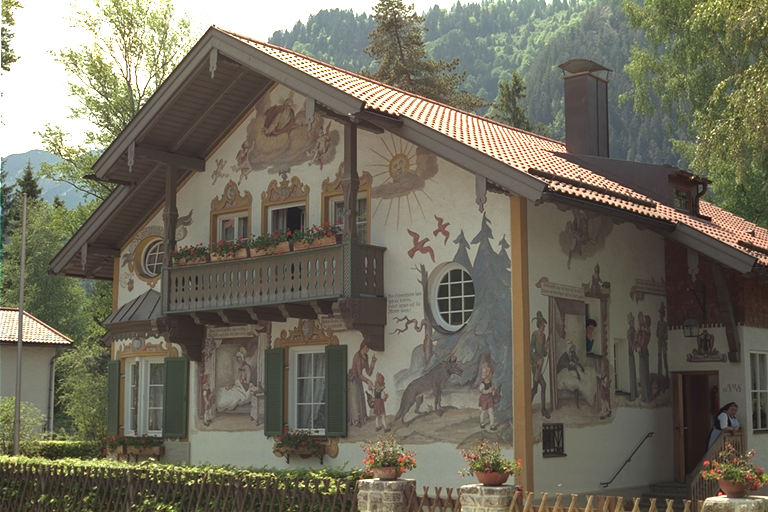}
        \subcaption{Ground Truth}
        \label{fig-RQob-Ori}
    \end{subfigure}\\
    \begin{subfigure}{0.48\columnwidth}
        \centering
        \includegraphics[width=\linewidth]{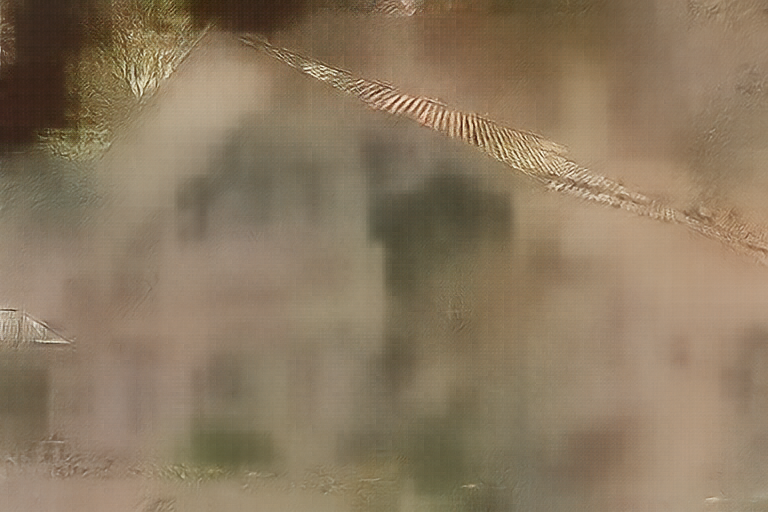}
        \subcaption{$\hat{i}_1 \neq i_1$, PSNR=10.854, MS-SSIM=0.000}
        \label{fig-RQob-step1}
    \end{subfigure}
    \hfill 
    \begin{subfigure}{0.48\columnwidth}
        \centering
        \includegraphics[width=\linewidth]{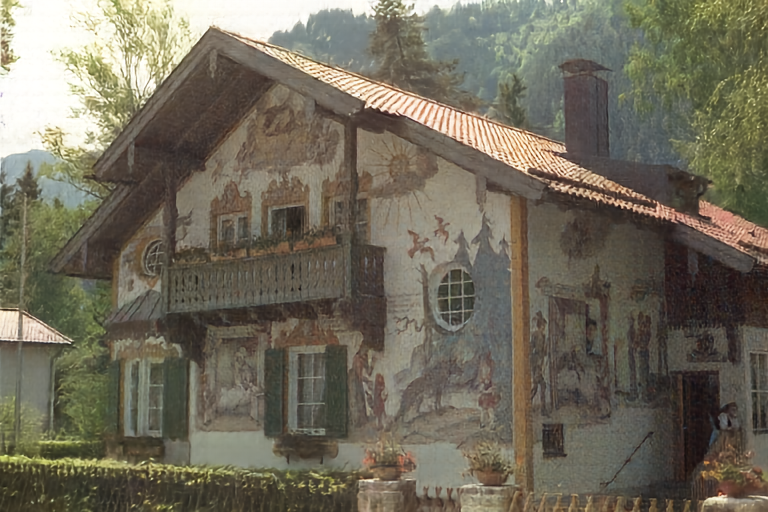}
        \subcaption{$\hat{i}_2 \neq i_2$, PSNR=25.342, MS-SSIM=0.929}
        \label{fig-RQob-step2}
    \end{subfigure}\\
    \begin{subfigure}{0.48\columnwidth}
        \centering
        \includegraphics[width=\linewidth]{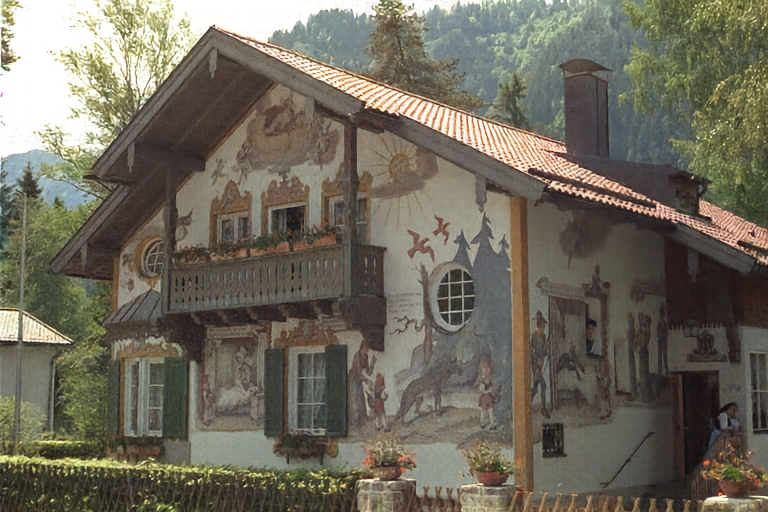}
        \subcaption{$\hat{i}_3 \neq i_3$, PSNR=27.182, MS-SSIM=0.964}
        \label{fig-RQob-step3}
    \end{subfigure}
    \hfill 
    \begin{subfigure}{0.48\columnwidth}
        \centering
        \includegraphics[width=\linewidth]{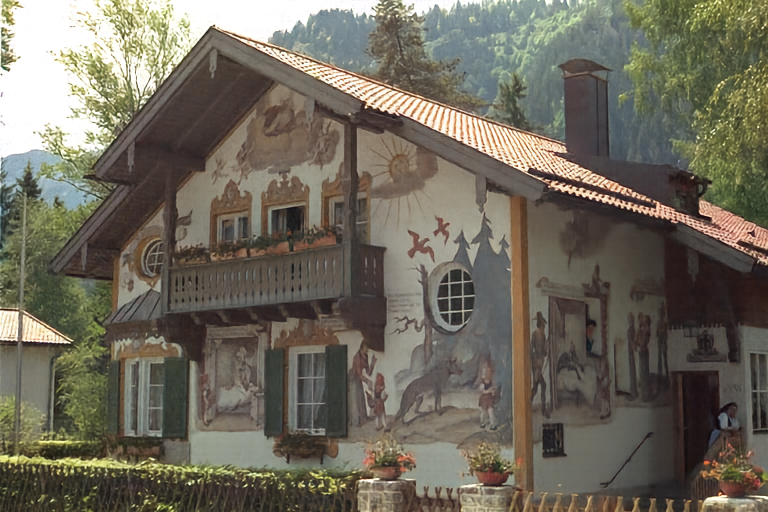}
        \subcaption{$\hat{i}_4 \neq i_4$, PSNR=27.842, MS-SSIM=0.970}
        \label{fig-RQob-step4}
    \end{subfigure}
    \caption{Visual comparison and corresponding quality metrics (PSNR, MS-SSIM) for the SSC system (configured with $N_q=4$) when transmission errors occur in the codeword indices from different steps of RQ. (a) Ground truth / Error-free. (b) Errors in the indices from Step 1. (c) Errors in the indices from Step 2. (d) Errors in the indices from Step 3. (e) Errors in the indices from Step 4.
    }
    \label{fig-RQob}
\end{figure}
\section{Semantic-Aware Stream Splitting Design} \label{Sec-Split}
As shown in \figref{fig-SSC_system}, the splitter divides the semantic stream $\mathbf{c}$, output by the residual quantizer $RQ(\cdot)$, into a  port index stream $\mathbf{c}_{pi}$ and a constellation symbol stream $\mathbf{c}_{sym}$.
This section presents the core design of our semantic-aware stream splitting scheme, which is motivated by observations of the trained SSC network and the analysis of the FA-IM transmission characteristics.

\begin{figure}[t]
    \centering
    \begin{subfigure}{0.49\columnwidth}
        \centering
        \includegraphics[width=\linewidth]{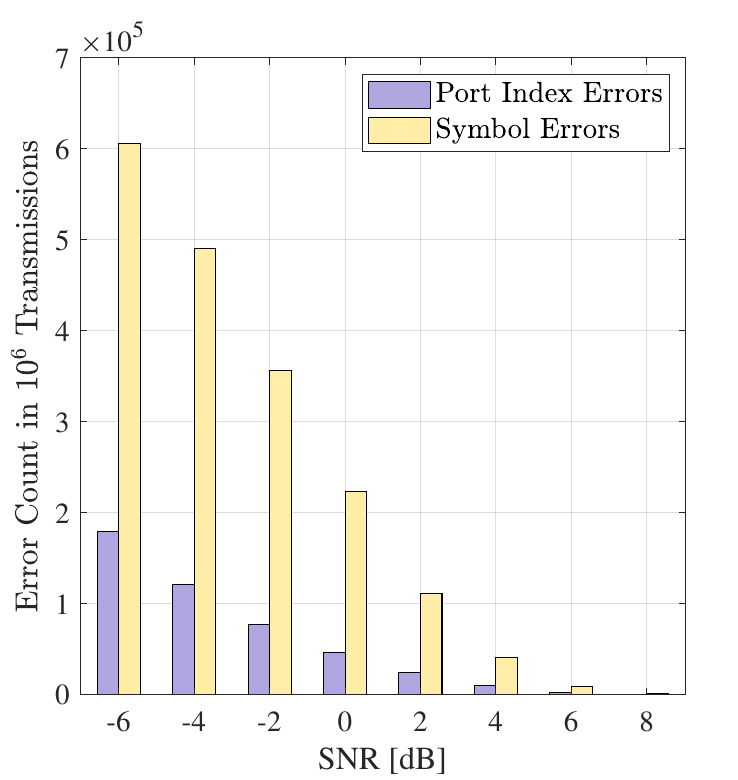} 
        \subcaption{$N_s=2, M=16$}
        \label{fig-FAIM_Ns2M16}
    \end{subfigure}
    \hfill 
    \begin{subfigure}{0.49\columnwidth}
        \centering
        \includegraphics[width=\linewidth]{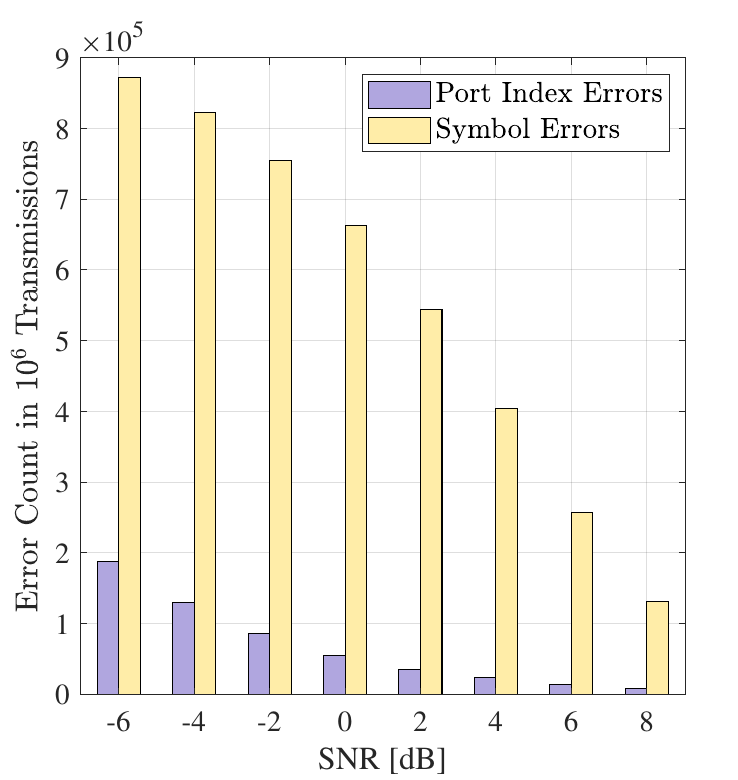}
        \subcaption{$N_s=2, M=64$}
        \label{fig-FAIM_Ns2M64}
    \end{subfigure}
    \\ 
    \begin{subfigure}{0.49\columnwidth}
        \centering
        \includegraphics[width=\linewidth]{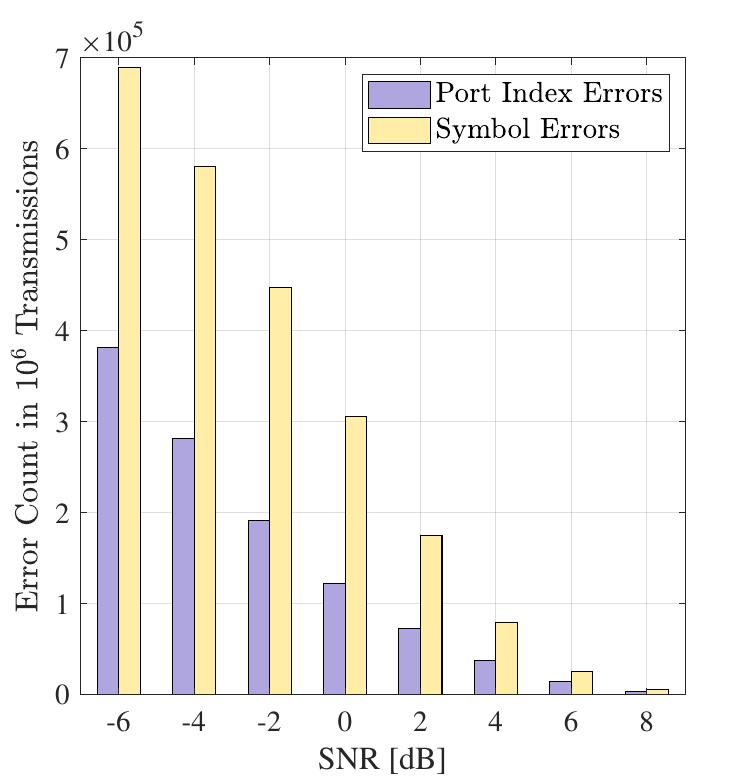}
        \subcaption{$N_s=4, M=16$}
        \label{fig-FAIM_Ns4M16}
    \end{subfigure}
    \hfill 
    \begin{subfigure}{0.49\columnwidth}
        \centering
        \includegraphics[width=\linewidth]{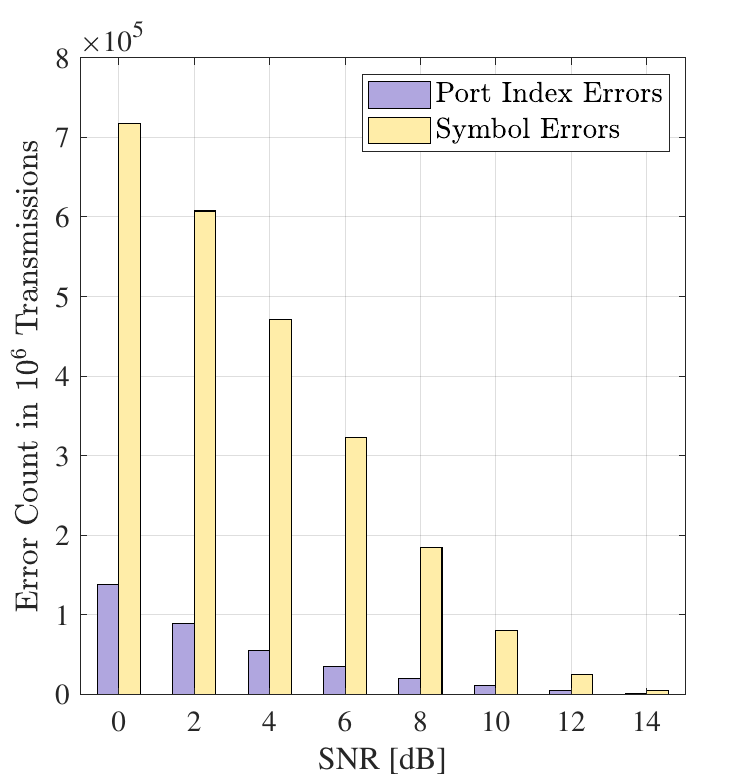}
        \subcaption{$N_s=4, M=64$}
        \label{fig-FAIM_Ns4M64}
    \end{subfigure}
    \caption{Error counts for port indices and symbols of the FA-IM scheme over $10^6$ transmission slots versus SNR. The fixed parameters are $W_t=2$, $N_p=16$, and $N_r=8$, while $N_s$ and $M$ vary as follows: (a) $N_s=2, M=16$; (b) $N_s=2, M=64$; (c) $N_s=4, M=16$; and (d) $N_s=4, M=64$.}
    \label{fig-FAIM_error}
\end{figure}
\subsection{Observations} \label{Sec-Split-Ob}
Recalling the operation of RQ in \eqref{eq-RQ} and the proposed commitment loss in \eqref{eq-loss_commit}, it follows that in our SSC system, for a given total number of quantization steps $N_q$, the quantization error progressively decreases as the step number $q$ increases. 
In other words, the smaller the value of $q$, the more semantic information is carried by the codeword $\mathbf{e}_{i_q}$ obtained at that step.
To validate this hypothesis, an experiment was conducted that an image was fed into the trained SSC network while the output of the residual quantizer was intentionally altered. 
Specifically, for a designated quantization step, its codeword indices were modified to incorrect values before being passed to the dequantizer. 
\figref{fig-RQob} visualizes the impact of index corruption at different quantization steps for $N_q=4$, presenting the reconstructed images alongside their corresponding peak signal-to-noise ratio (PSNR) and multi-scale structural similarity (MS-SSIM) scores. 
As expected, transmission errors in the indices from quantization Step 1 leads to severe and visually perceptible distortions in the reconstructed image, as shown in \figref{fig-RQob-step1}. 
As the corrupted quantization step is progressively delayed, i.e., for larger $q$, the reconstruction quality steadily improves. 
Therefore, the results in \figref{fig-RQob} provide compelling evidence for our hypothesis that earlier quantization steps contain more significant semantic information.

From another perspective, the information stream entering the FA-IM modulator is split into two separate paths. 
One path is conveyed through the selection of activated port indices, while the other is transmitted via the constellation symbols. 
A natural consequence of this splitting is that the two streams are subjected to different fading effects.
To visually demonstrate this disparity, we simulated the FA-IM transmission scheme and counted the number of errors for both the port indices and the constellation symbols over $10^6$ transmission slots (one port index and one constellation symbol are transmitted per slot). 
The results are presented in \figref{fig-FAIM_error}, where the fixed system parameters were set to $W_t=2$, $N_p=16$, and $N_r=8$, while different combinations of the number of active ports $N_s \in \{2, 4\}$ and modulation order $M \in \{16, 64\}$ were evaluated.
As can be observed, under current configurations, the number of transmission errors for port indices is lower than that for constellation symbols, particularly in the low SNR region.
This observation validates that the two data streams undergo distinct fading dynamics, consequently exhibiting different error performance.

\begin{algorithm}[t]
\caption{Semantic-aware Stream Splitting Scheme}
\label{alg-splitting} 
\renewcommand{\algorithmicrequire}{\textbf{Input:}}
\renewcommand{\algorithmicensure}{\textbf{Output:}}
\begin{algorithmic}[1]
    \REQUIRE Semantic information sequence $\mathbf{f}$; FA-IM transmission parameters: $N_s$, $M$, $W_t$, $N_p$, $N_r$.
    \ENSURE Port index stream $\mathbf{c}_{pi}$ and symbol stream $\mathbf{c}_{sym}$.
    
    \STATE // \textit{Robustness Analysis}
    \STATE Execute Monte Carlo simulations for FA-IM configured with $W_t, N_p, N_r, N_s, M$;
    \STATE Obtain error statistics $e_{pi}$ and $e_{sym}$;
    \STATE // \textit{Stream Splitting}
    \STATE Calculate the allocated lengths $L_{pi}$ and $L_{sym}$ based on \eqref{eq-split2_point};
    
    \IF{$e_{pi} < e_{sym}$}
        \STATE // \textit{Assign important semantic information to $\mathbf{c}_{pi}$}
        \STATE $\mathbf{c}_{pi} \leftarrow [f_1, \ldots, f_{L_{pi}}]$;
        \STATE $\mathbf{c}_{sym} \leftarrow [f_{L_{pi}+1}, \ldots, f_{L_u N_q}]$;
    \ELSE
        \STATE // \textit{Assign important semantic information to $\mathbf{c}_{sym}$}
        \STATE $\mathbf{c}_{sym} \leftarrow [f_1, \ldots, f_{L_{sym}}]$;
        \STATE $\mathbf{c}_{pi} \leftarrow [f_{L_{sym}+1}, \ldots, f_{L_u N_q}]$;
    \ENDIF
    
    \RETURN $\mathbf{c}_{pi}, \mathbf{c}_{sym}$.
\end{algorithmic} 
\end{algorithm}
\subsection{Splitting Design} \label{Sec-Split-Design}
Based on the observations and inferences from \figref{fig-RQob} and \figref{fig-FAIM_error}, we propose a novel semantic-aware stream splitting scheme for the SSC system. 
The core idea is to leverage the observed asymmetry in both semantic importance and transmission reliability.
The more semantically important codeword indices from early quantization steps are allocated to the more robust data stream for transmission. 
Conversely, the indices from the later, less significant quantization steps are transmitted via the other data stream.

More specifically, to prioritize the semantic information flow based on its importance, the splitter first reshapes the input codeword index sequence $\mathbf{c}$ of length $L_u N_q$ into an $L_u \times N_q$ matrix $\mathbf{A}$, which is represented as $\mathbf{A}=[\mathbf{c}_1; \mathbf{c}_2; \ldots; \mathbf{c}_{L_u}]=[\mathbf{a}_1, \mathbf{a}_2, \ldots, \mathbf{a}_{N_q}]$, with each column $\mathbf{a}_q \in \{1, 2, \ldots, N_e\}^{L_u}$.
The matrix $\mathbf{A}$ is then flattened into a new sequence $\mathbf{f}=[\mathbf{a}_1^T, \mathbf{a}_2^T, \ldots, \mathbf{a}_{N_q}^T]=[f_1, f_2, \ldots, f_{L_u N_q}]$. 
Through this operation, the sequence $\mathbf{f}$ is successfully arranged in order of semantic importance, such that elements positioned earlier in the sequence are semantically more significant.

According to the operational principles of FA-IM, the respective lengths of the port index stream, $L_{pi}$, and the constellation symbol stream, $L_{sym}$, are calculated as 
\begin{equation}
\begin{aligned}
\label{eq-split2_point}
L_{pi} &= \text{round} \left( \frac{m_1}{m_1 + m_2} \cdot L_u N_q \right),  \\
L_{sym} &= L_u N_q - L_{pi}
\end{aligned}
\end{equation}%
where $\text{round}(\cdot)$ denotes the rounding operation to the nearest integer.
Next, to identify the more robust data stream of FA-IM transmission, Monte Carlo simulations are conducted to obtain the error statistics for port indices and constellation symbols, denoted as $e_{pi}$ and $e_{sym}$, respectively.
Based on the comparison between $e_{pi}$ and $e_{sym}$, the leading indices in sequence $\mathbf{f}$, which carry the most critical semantic content, are allocated to the data stream exhibiting superior error resilience. 
Conversely, the remaining fine-grained refinement indices are allocated to the other stream with higher error statistics.
The proposed semantic-aware stream splitting scheme is summarized in Algorithm \ref{alg-splitting}.
This strategy ensures that the essential semantic features are protected by the more reliable transmission dimension in FA-IM, thereby maximizing reconstruction fidelity.
In the subsequent simulation results, we establish a baseline that simply treats the physical IM transmission as a blind pipe without this semantic-aware stream splitting scheme (i.e., the ``SSC w/o SS'' baseline). 
Comparing this to the complete SSC scheme, the simulation results highlight the significant performance gains brought by semantic splitting, validating that this cross-layer co-design is essential to fully unleash the joint potential of semantic communications and physical-layer IM.

It is important to emphasize that the Monte Carlo simulation outlined in Algorithm \ref{alg-splitting} is executed strictly offline as a one-time pre-configuration step during the initial system design phase. 
In practical deployment, the stream-splitting decision is governed solely by the relative order of the two transmission streams, i.e., whether $e_{pi}$ or $e_{sym}$ is larger, irrespective of their absolute, instantaneous error values.
Crucially, as validated by the error count results in \figref{fig-FAIM_error}, this relative relationship is a deterministic structural property of the FA-IM framework, uniquely determined by static system parameters such as the active port count $N_s$ and the modulation order $M$. 
Since the relative order of these errors is invariant to fast-varying fading and SNR fluctuations, the stream-splitting mapping can be pre-determined for a given hardware configuration, thereby eliminating online computational overhead and real-time scheme updates.

It is worth noting that the above design philosophy is conceptually related to the classical principle of UEP. Our contribution lies in instantiating this well-established principle in the previously unexamined context of jointly designed digital semantic quantization and physical-layer IM. 
Specifically, existing digital semantic communication frameworks typically treat the physical layer as a homogeneous, transparent pipe, while conventional IM designs treat the dual-stream physical asymmetry solely as a means to transmit unstructured, raw bits. 
By identifying and exploiting the correspondence between the source-side hierarchical significance of multi-step RQ and the physical-layer unequal error statistics of FA-IM, our scheme establishes a concrete cross-layer link between these two previously separately treated design spaces.
The core concept of our proposed scheme can be readily extended to other progressive or hierarchical digital semantic technologies and a wide array of IM variants that exhibit dual-stream physical reliability asymmetry.

\section{Simulation Results}  \label{Sec-Simulation}
In this section, the simulation results are presented to evaluate the performance of the proposed SSC system. 

\begin{figure*}[t]
    \centering
    \begin{subfigure}{0.32\textwidth}
        \centering
        \includegraphics[width=0.9\linewidth]{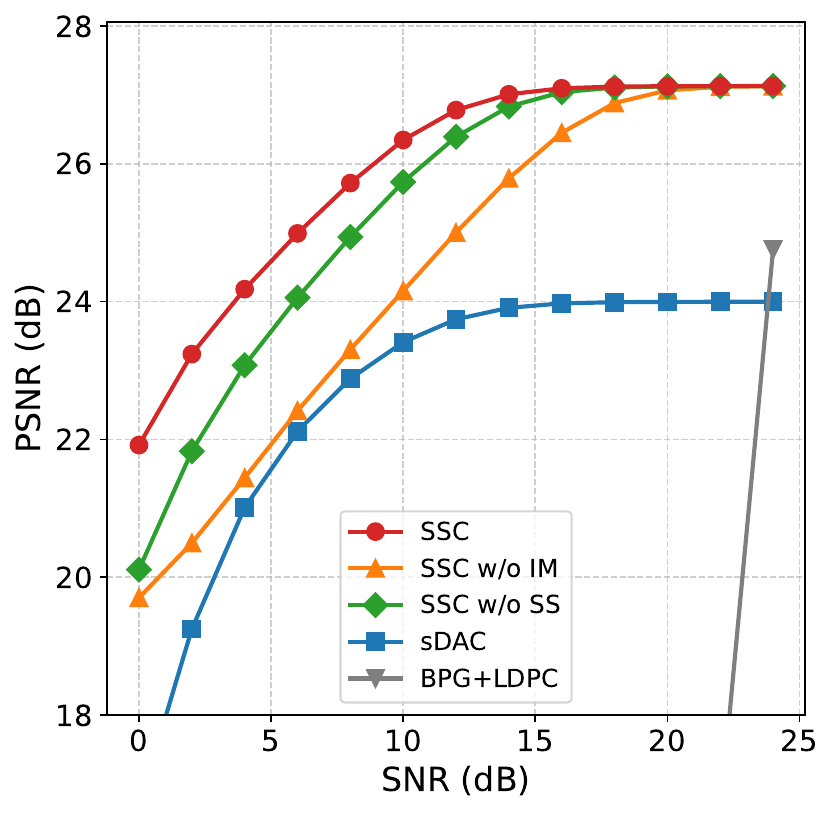}
        \subcaption{PSNR at 2.0 bpp}
        \label{fig-bpp2_psnr}
    \end{subfigure}
    \hfill
    \begin{subfigure}{0.32\textwidth}
        \centering
        \includegraphics[width=0.9\linewidth]{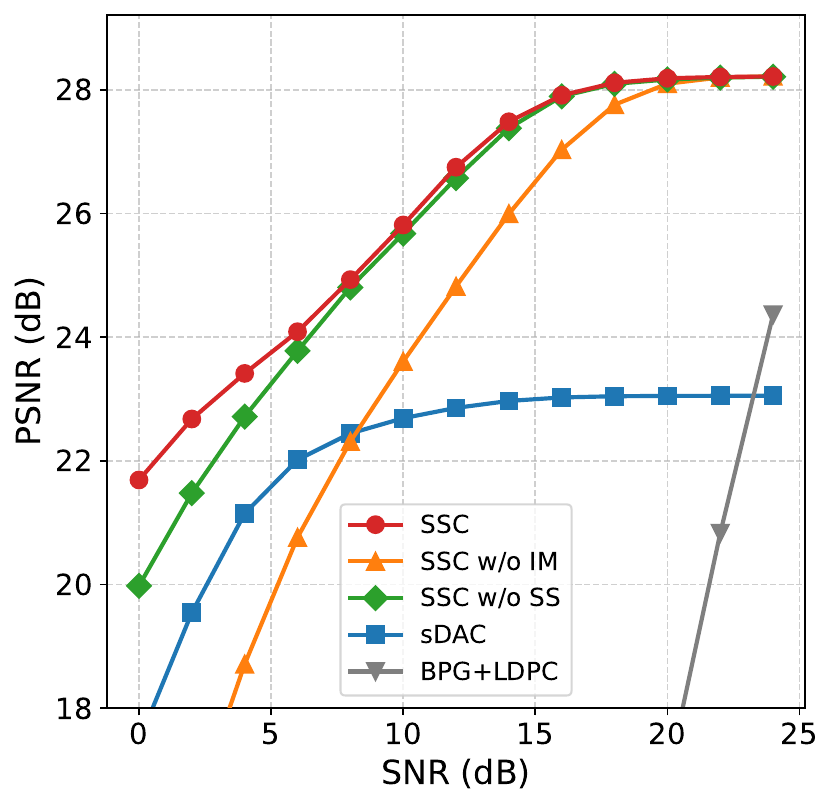}
        \subcaption{PSNR at 4.0 bpp}
        \label{fig-bpp4_psnr}
    \end{subfigure}
    \hfill
    \begin{subfigure}{0.32\textwidth}
        \centering
        \includegraphics[width=0.9\linewidth]{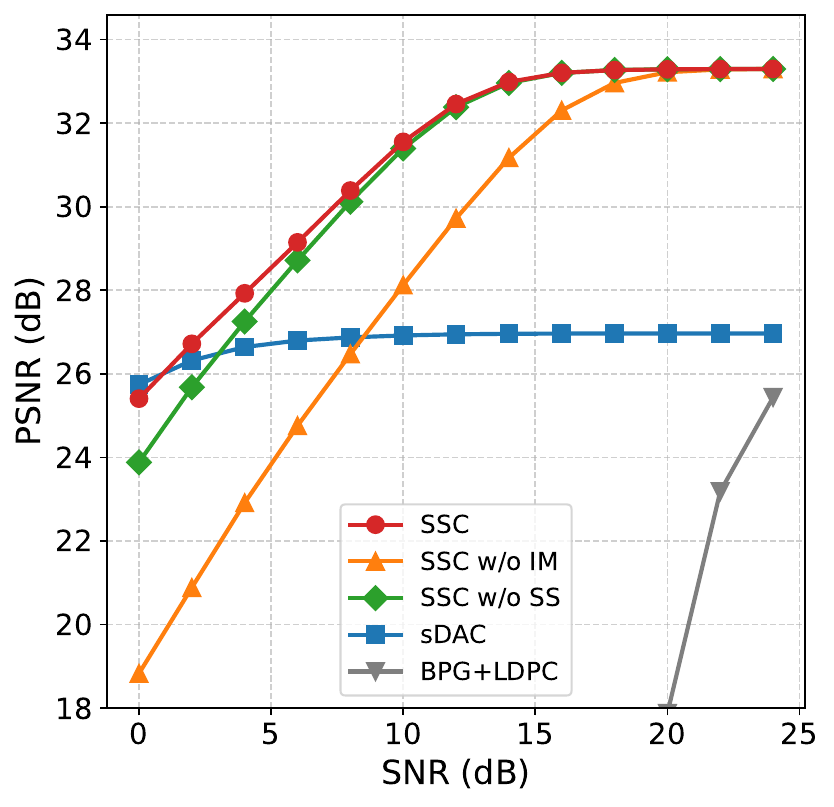}
        \subcaption{PSNR at 6.0 bpp}
        \label{fig-bpp6_psnr}
    \end{subfigure}\\
    \begin{subfigure}{0.32\textwidth}
        \centering
        \includegraphics[width=0.9\linewidth]{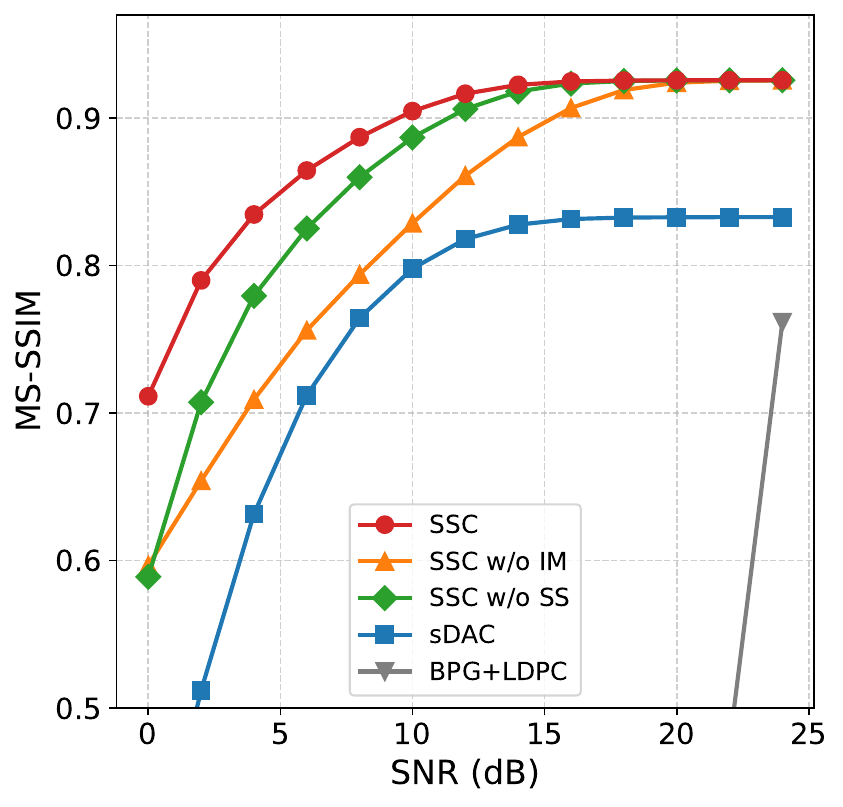}
        \subcaption{MS-SSIM at 2.0 bpp}
        \label{fig-bpp2_ms_ssim}
    \end{subfigure}
    \hfill
    \begin{subfigure}{0.32\textwidth}
        \centering
        \includegraphics[width=0.9\linewidth]{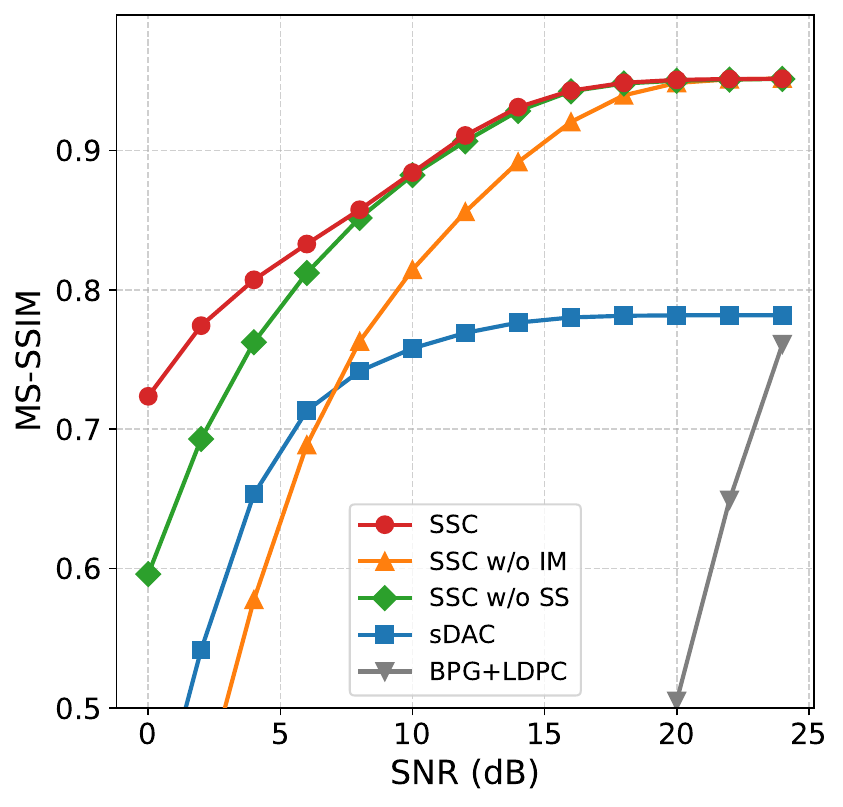}
        \subcaption{MS-SSIM at 4.0 bpp}
        \label{fig-bpp4_ms_ssim}
    \end{subfigure}
    \hfill
    \begin{subfigure}{0.32\textwidth}
        \centering
        \includegraphics[width=0.9\linewidth]{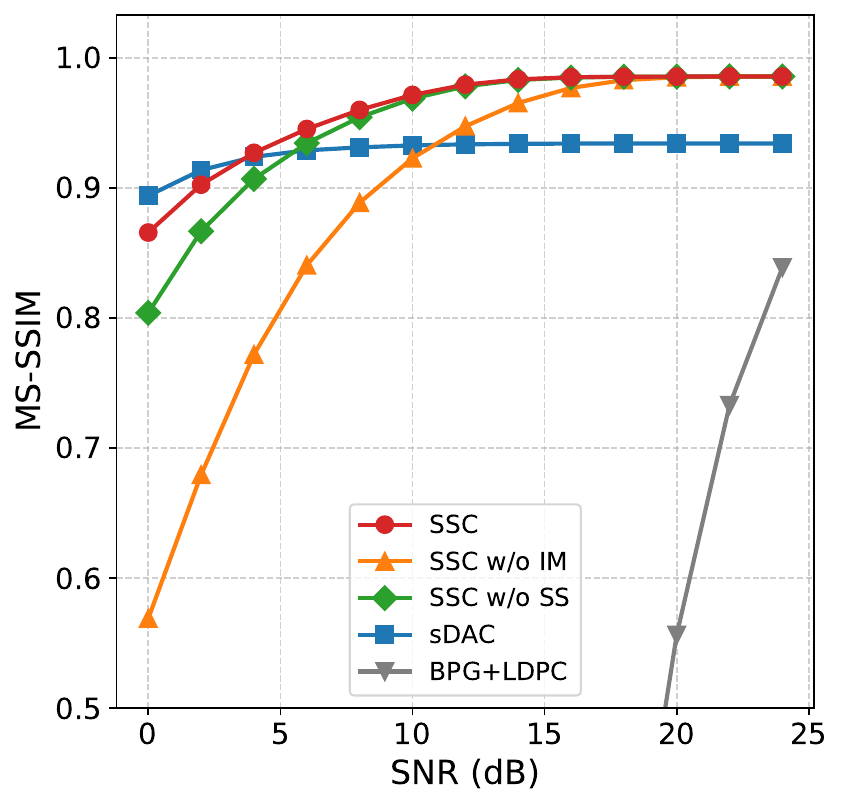}
        \subcaption{MS-SSIM at 6.0 bpp}
        \label{fig-bpp6_ms_ssim}
    \end{subfigure}

    \caption{Performance comparison of PSNR and MS-SSIM between the proposed SSC system and benchmarks across various bit rates. (a) PSNR at 2.0 bpp ($C_z=32$). (b) PSNR at 4.0 bpp ($C_z=64$). (c) PSNR at 6.0 bpp ($C_z=96$). (d) MS-SSIM at 2.0 bpp ($C_z=32$). (e) MS-SSIM at 4.0 bpp ($C_z=64$). (f) MS-SSIM at 6.0 bpp ($C_z=96$).}
    \label{fig-bpp_psnr_ms_ssim}
\end{figure*}
\subsection{Experimental Setups} \label{Sec-Simulation-Setup}
\subsubsection{Implementation Details}
Focusing on the image reconstruction task, we employ the `small' version of the widely adopted SwinJSCC \cite{ref-SwinJSCC}, which is based on the Swin Transformer \cite{ref-SwinT}, to serve as the JSCC backbone of our SSC system.
The model is trained on the DIV2K dataset \cite{ref-DIV2K} and evaluated on the Kodak24 dataset \cite{ref-Kodak24}.
Specifically, the SSC system is trained following the three-stage strategy illustrated in \figref{fig-SSC_training}, with the batch size set to 16.
For Stage 0, the JSCC backbone is trained for a total of $5 \times 10^5$ iterations. 
We employ the Adam optimizer with a learning rate of $3 \times 10^{-5}$, and betas set to $(0.9, 0.99)$. 
To ensure training stability, the EMA decay rate is set to 0.999.
The learning rate follows a multi-step decay schedule with milestones at iterations $\{250\text{k}, 400\text{k}, 450\text{k}, 475\text{k}\}$ and a decay factor 0.5. 
Alternatively, this stage can be bypassed by loading pre-trained SwinJSCC weights to accelerate the process.
For Stage 1 and Stage 2, the training process spans a total of $2 \times 10^5$ iterations. 
Crucially, the first $1 \times 10^5$ iterations correspond to Stage 1, during which the backbone is frozen to facilitate the stable initialization of the RQ module. 
Subsequently, for the remaining iterations (Stage 2), the entire network is unfrozen to enable global fine-tuning.
The learning rate is decayed by a factor of $0.5$ at iterations $[110\text{k}, 140\text{k}, 180\text{k}]$. 
During each forward propagation, the bit-flip probability $p$ of the BSC is randomly sampled from the predefined set $\{1.31 \times 10^{-1}, 5.63 \times 10^{-2}, 1.25 \times 10^{-2}, 7.73 \times 10^{-4}, 3.87 \times 10^{-6}, 1.33 \times 10^{-10}, 2.27 \times 10^{-19}\}$, which is calculated from uniformly spaced SNR values.
The weight of the commitment loss is set to $\beta = 0.25$, and the decay factor for the codebook EMA update is set to $\gamma = 0.99$.
Unless otherwise specified, the default parameters of the SSC network are set as follows: the SwinJSCC encoder generates the semantic feature $\mathbf{Z}$ with a length of $L_z=256$ and a channel dimension of $C_z =32$; the codebook $\mathcal{B}$ is configured with a codeword dimension of $d_e=4$ and a codebook size of $N_e=16$, while the number of RQ steps is $N_q=4$.
Conducted on two NVIDIA RTX 4090 GPUs (24 GB), the training process requires approximately 2 days for Stage 0 and less than 1 day for the subsequent Stages 1 and 2.
To evaluate the reconstruction quality, we employ the widely used pixel-wise PSNR and the perceptual MS-SSIM. 
Furthermore, during the evaluation phase, the actual FA-IM channel is implemented. 
Unless stated otherwise, the default channel and modulation parameters are set to: $L_p=10$, $W_t=2$, $N_p=16$, $N_s=4$, $M=64$, and $N_r=8$.
All experiments are implemented using PyTorch.

\subsubsection{Benchmarks}
To demonstrate the superiority of the proposed SSC system, we conduct comparative experiments against the following benchmarks.
\begin{itemize}
    \item \textbf{sDAC}: sDAC is a plug-and-play digitization module \cite{ref-sDAC}. 
    This baseline utilizes the same SwinJSCC backbone as the SSC system, and its implementation details are in strict accordance with its original design. 
    Specifically, we adhere faithfully to its network architecture, VQ mechanism, and training strategy pipeline. 
    To align the transmission rate under identical system settings, the sDAC codebook size is configured to act as the exact equivalent representation capacity to our RQ. 
    For the modulation interface, the quantized binary bits generated by sDAC are directly mapped to the identical FA-IM modulator (incorporating both port indices and constellation symbols) to ensure that sDAC is evaluated over the exact same physical-layer carrier and channel environment as the proposed SSC. 
    \item \textbf{SSC w/o IM}: This scheme employs conventional QAM, where the FA activates the specific port with the maximum channel gain for symbol transmission \cite{ref-FAS}. 
    For a fair comparison, the constellation size is set to $M=256$ to align the SE with the proposed FA-IM counterpart, as defined in \eqref{eq-SE_SSC}.
    \item \textbf{SSC w/o SS}: This scheme operates without the \textbf{S}emantic stream \textbf{S}plitting (SS) proposed in Section \ref{Sec-Split}. 
    Instead, the index sequence $\mathbf{c}$ is split randomly and fed into the FA-IM modulator, disregarding the semantic significance of the codewords.
    \item \textbf{BPG+LDPC}: This benchmark adopts the traditional separation-based source-channel coding framework to produce a bitstream for FA-IM transmission. 
    Specifically, it utilizes the better portable graphics (BPG) codec \cite{ref-BPG} for source coding and the IEEE 802.11 (WiFi) standard low-density parity-check (LDPC) codes \cite{ref-LDPC} for channel coding, configured with a block length of 1944.
\end{itemize}

\subsection{Comparative Performance across Compression Rates} \label{Sec-Simulation-bpp}
First, we assess the performance of the proposed SSC system and the considered benchmarks across different compression rates. 
To ensure fairness, all schemes are evaluated under equivalent bits-per-pixel (bpp) levels. 
The bpp of the proposed SSC system is calculated as 
\begin{equation}
\label{eq-bpp}
\text{bpp} = \frac{L_z C_z N_q \log_2 N_e}{H_i W_i},  
\end{equation}%
where $H_i$ and $W_i$ denote the height and width of the input image to the JSCC encoder, respectively (with $H_i=W_i=256$ in the adopted SwinJSCC). 
For the proposed SSC system, the compression rate is varied by adjusting the channel dimension of the semantic features output by the SwinJSCC encoder, specifically $C_z \in \{32, 64, 96\}$.
For the VQ-based ``\textbf{sDAC}'' benchmark, we maintain a consistent $C_z$ and scale its codebook size $N_e^\prime$ to match the bpp of the proposed SSC system.
Following the relationship $N_e^\prime = N_e^{N_q}$, the codebook size is set to $N_e^\prime = 65536$.
Collectively, these configurations enable both systems to span standardized compression rates of $2.0$, $4.0$, and $6.0$ bpp.
Accordingly, \figref{fig-bpp_psnr_ms_ssim} presents the PSNR and MS-SSIM performance versus SNR, respectively.
Several key observations can be drawn from these results:

\begin{table}[t]
\centering
\caption{Complexity and Convergence Comparison between Proposed SSC and sDAC}
\label{tab-complexity-comparison}
\resizebox{\linewidth}{!}{
\begin{tabular}{|l|c|c|}
\hline
\textbf{Metric} & \textbf{sDAC\cite{ref-sDAC}} & \textbf{Proposed SSC} \\ \hline
Quantization Scheme & VQ & RQ \\ \hline
Quantization Steps $N_q$ & 1 & 4 \\ \hline
Codebook Size ($N'_e \& N_e$) & 65536 & 16 \\ \hline
Search Complexity (FLOPs) & $\mathcal{O}(N'_e \cdot d_e) = 262144$ & $\mathcal{O}(N_q \cdot N_e \cdot d_e) = 256$ \\ \hline
Convergence Time (Stages 1 \& 2) & $33$ hours & $14$ hours \\ \hline
\end{tabular}
} 
\end{table}
1) Compared to the sDAC benchmark, the proposed SSC system achieves higher reconstruction fidelity across almost the entire SNR range. 
While sDAC occasionally demonstrates competitive performance at extremely low SNRs in high-bpp scenarios (e.g., \figref{fig-bpp6_psnr} and \figref{fig-bpp6_ms_ssim}), it quickly hits a performance ceiling and saturates as SNR increases. 
Notably, to maintain the same bpp as the proposed SSC system, the sDAC benchmark incurs prohibitive computational and storage overheads for quantization and codebook maintenance, as quantitatively detailed in Table \ref{tab-complexity-comparison}. 
Specifically, compared to our proposed SSC configured with $N_q=4$ and $N_e=16$, the sDAC benchmark requires 4096 times the storage space. 
Furthermore, standard VQ in sDAC demands calculating the Euclidean distance to all 65536 codewords during the nearest-neighbor search, resulting in a search complexity of $\mathcal{O}(N'_e \cdot d_e) = 262144$ FLOPs per latent vector. 
In contrast, the proposed RQ in SSC reduces this to $\mathcal{O}(N_q \cdot N_e \cdot d_e) = 256$ FLOPs per latent vector, achieving a 1024-fold reduction in nearest-neighbor search complexity. 
These efficiency advantages also significantly enhance training stability and convergence. 
Under identical hardware conditions and using the same pre-trained Stage 0 JSCC backbone, the subsequent training of sDAC requires 33 hours to reach a stable reconstruction state, whereas our proposed SSC system converges in only 14 hours.
The single-shot quantization nature of VQ constrains its representation capability, and its performance is further degraded by the well-known codebook collapse issue as the codebook size increases.
Conversely, the proposed SSC, by leveraging RQ with more quantization steps, effectively expands the equivalent codebook size exponentially without incurring the prohibitive storage overhead.
This efficiency allows the SSC system to break the performance bottleneck, thereby achieving a much higher upper bound in both PSNR and MS-SSIM.

2) The pronounced performance gap between the proposed SSC system and the ``SSC w/o IM'' benchmark highlights the pivotal role of integrating index modulation. 
Operating under identical SE constraints, the ``SSC w/o IM'' curve, which relies solely on constellation modulation, consistently lags behind the SSC system. 
With the integration of IM, the SSC system achieves substantial gains in the low-SNR regime and reaches its performance upper bound more rapidly.
This evidence substantiates that exploiting the spatial domain indices of the FA to convey semantic information effectively bolsters the system's resilience against channel impairments.

3) Comparing the red ``SSC'' curves with the green ``SSC w/o SS'' curves, we validate the effectiveness of the proposed semantic-aware stream splitting design. 
The SSC system yields higher PSNR and MS-SSIM scores than its random-splitting counterpart, with the performance advantage being most pronounced in the low SNR regions. 
This improvement is attributed to the unequal error protection mechanism inherent in our design: by mapping the coarse-grained, semantically critical quantization indices (from early RQ steps) to the highly reliable data stream of the FA-IM modulator, the system ensures that the most fundamental semantic features are preserved even under adverse channel conditions.

4) As observed in all subplots, the traditional separation-based ``BPG+LDPC'' scheme exhibits a severe cliff effect. 
Due to the characteristics of digital coding, its performance drops precipitously to unacceptable levels when the channel capacity falls below the transmission rate. 
In sharp contrast, deep learning-based JSCC solutions, including the proposed SSC system, demonstrate remarkable graceful degradation. 
Even in low SNR regimes, the SSC system maintains intelligible image reconstruction quality, validating the inherent robustness of the JSCC paradigm and the proposed transmission strategy.

In summary, the proposed SSC system effectively combines the robustness of JSCC, the high fidelity of RQ, and the spatial efficiency of FA-IM, demonstrating superior and comprehensive performance compared to all baselines across various bandwidth and channel conditions.

\begin{figure}[t]
    \centering
    \begin{subfigure}{0.7\columnwidth}
        \centering
        \includegraphics[width=\linewidth]{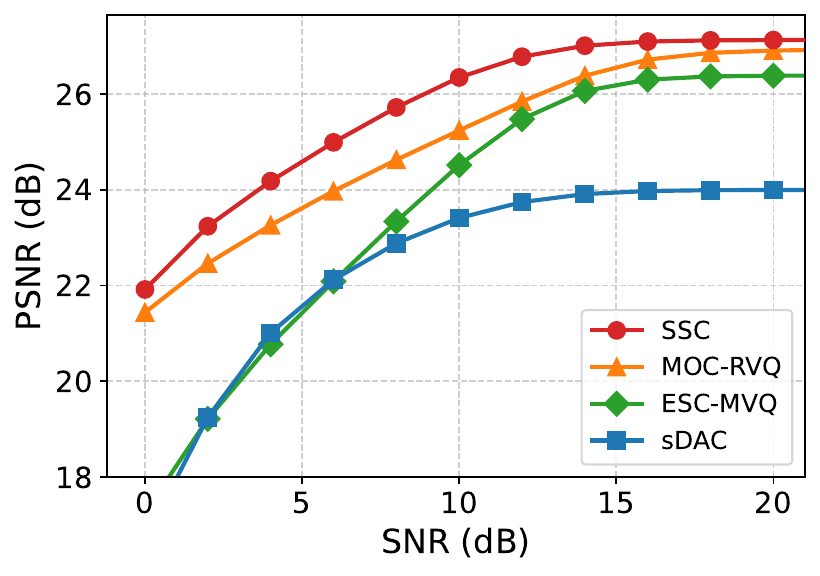}
        \subcaption{PSNR}
        \label{fig-baseline_psnr}
    \end{subfigure}\\
    \begin{subfigure}{0.7\columnwidth}
        \centering
        \includegraphics[width=\linewidth]{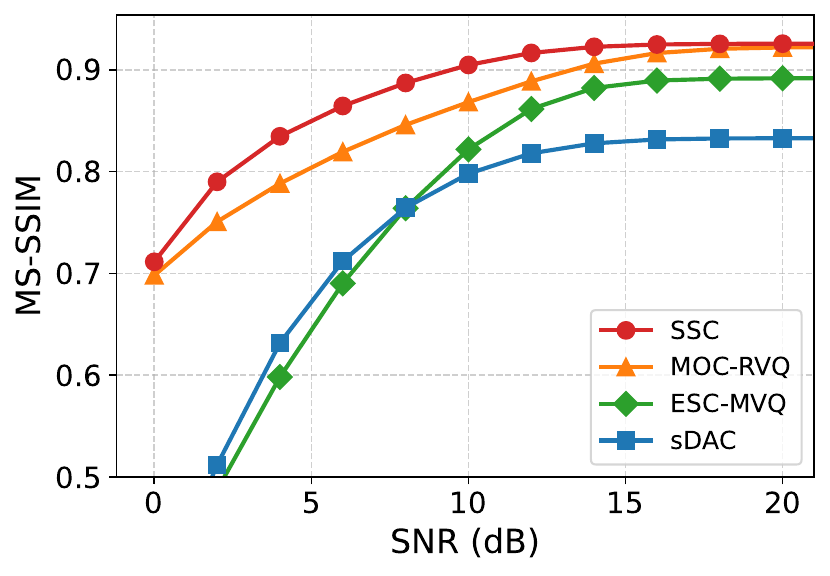}
        \subcaption{MS-SSIM}
        \label{fig-baseline_ms_ssim}
    \end{subfigure}
    \caption{Performance comparisons between the proposed SSC system and more SOTA baselines. (a) PSNR. (b) MS-SSIM.}
    \label{fig-baseline_psnr_ms_ssim}
\end{figure}
\begin{figure}[t]
    \centering
    \begin{subfigure}{0.49\columnwidth}
        \centering
        \includegraphics[width=\linewidth]{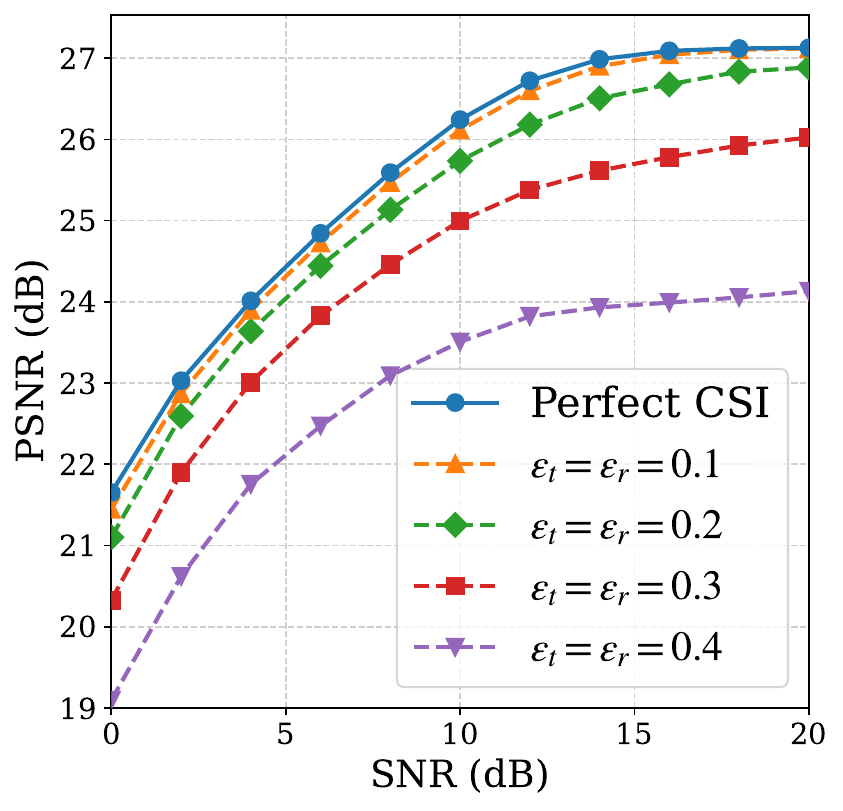}
        \subcaption{PSNR under joint errors}
        \label{fig-csi_tr_PSNR}
    \end{subfigure}
    \hfill 
    \begin{subfigure}{0.49\columnwidth}
        \centering
        \includegraphics[width=\linewidth]{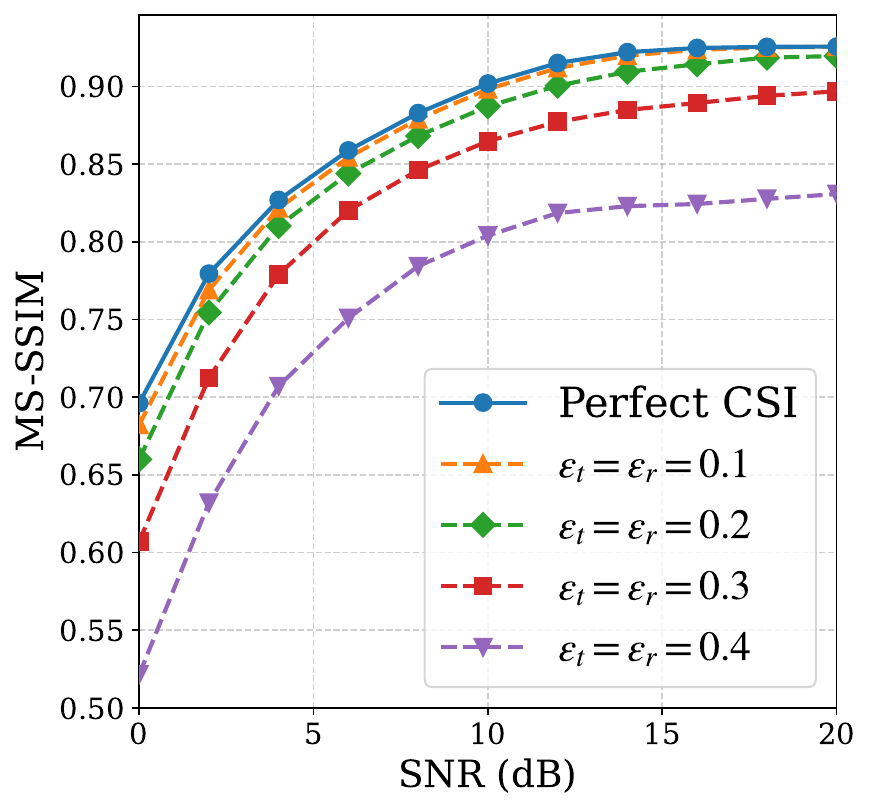}
        \subcaption{MS-SSIM under joint errors}
        \label{fig-csi_tr_MS_SSIM}
    \end{subfigure}
    \\
    \begin{subfigure}{0.49\columnwidth}
        \centering
        \includegraphics[width=\linewidth]{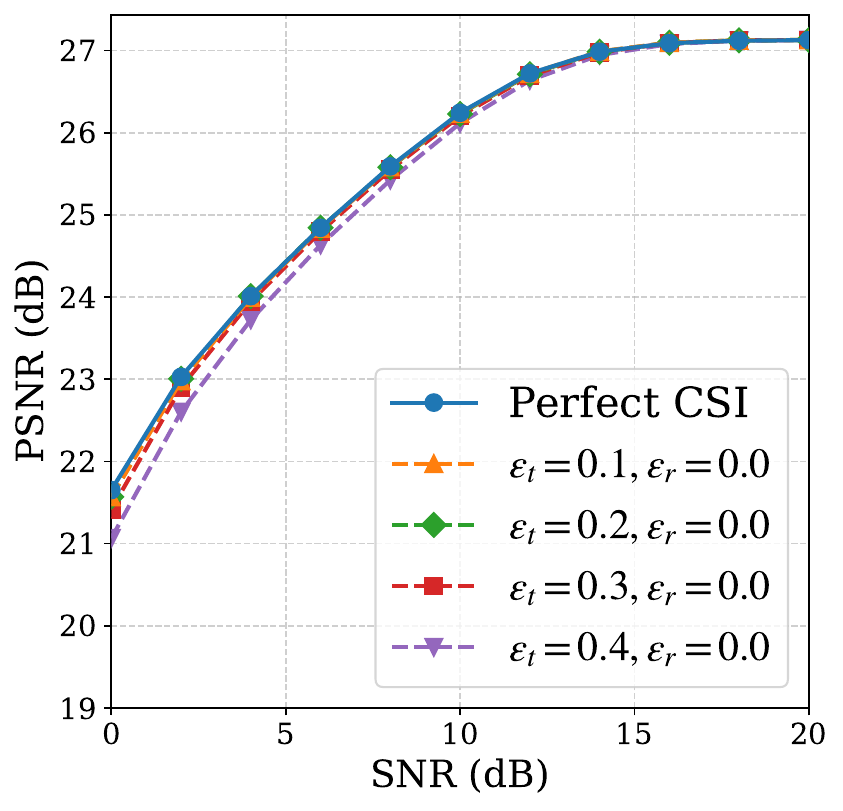}
        \subcaption{PSNR under TX-side errors ($\varepsilon_r = 0$)}
        \label{fig-csi_tx_PSNR}
    \end{subfigure}
    \hfill 
    \begin{subfigure}{0.49\columnwidth}
        \centering
        \includegraphics[width=\linewidth]{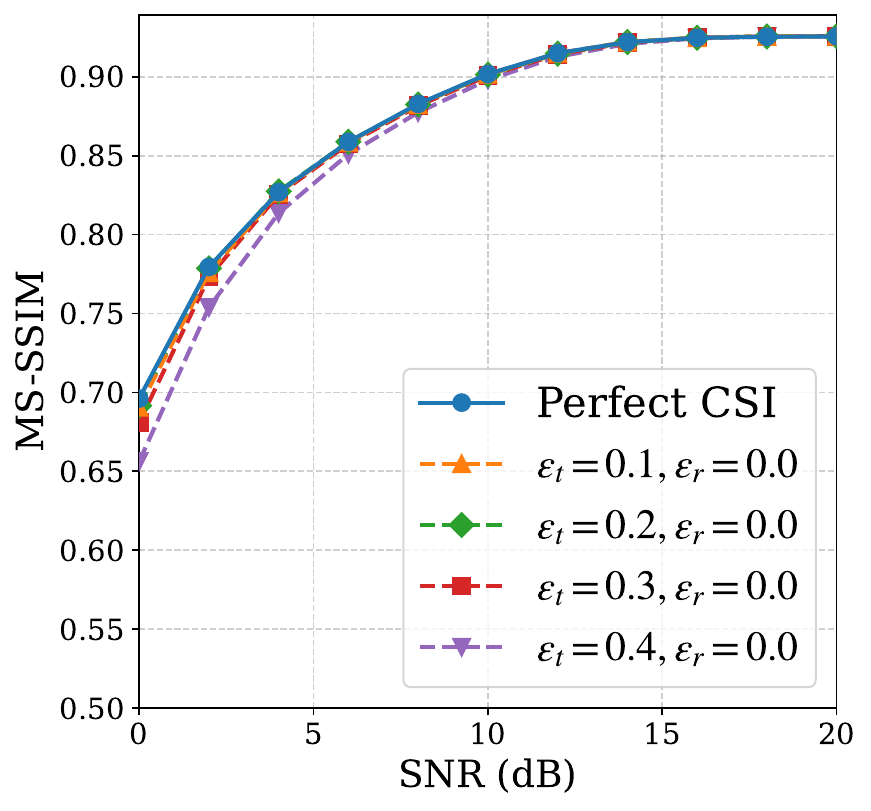}
        \subcaption{MS-SSIM under TX-side errors ($\varepsilon_r = 0$)}
        \label{fig-csi_tx_MS_SSIM}
    \end{subfigure}
    \\
    \begin{subfigure}{0.49\columnwidth}
        \centering
        \includegraphics[width=\linewidth]{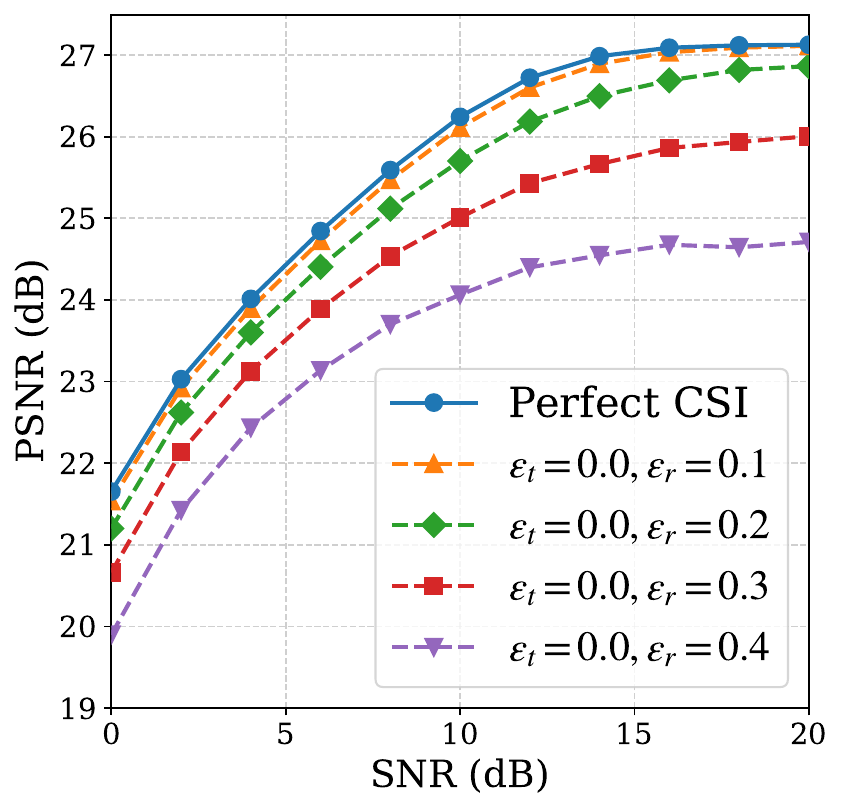}
        \subcaption{PSNR under RX-side errors ($\varepsilon_t = 0$)}
        \label{fig-csi_rx_PSNR}
    \end{subfigure}
    \hfill 
    \begin{subfigure}{0.49\columnwidth}
        \centering
        \includegraphics[width=\linewidth]{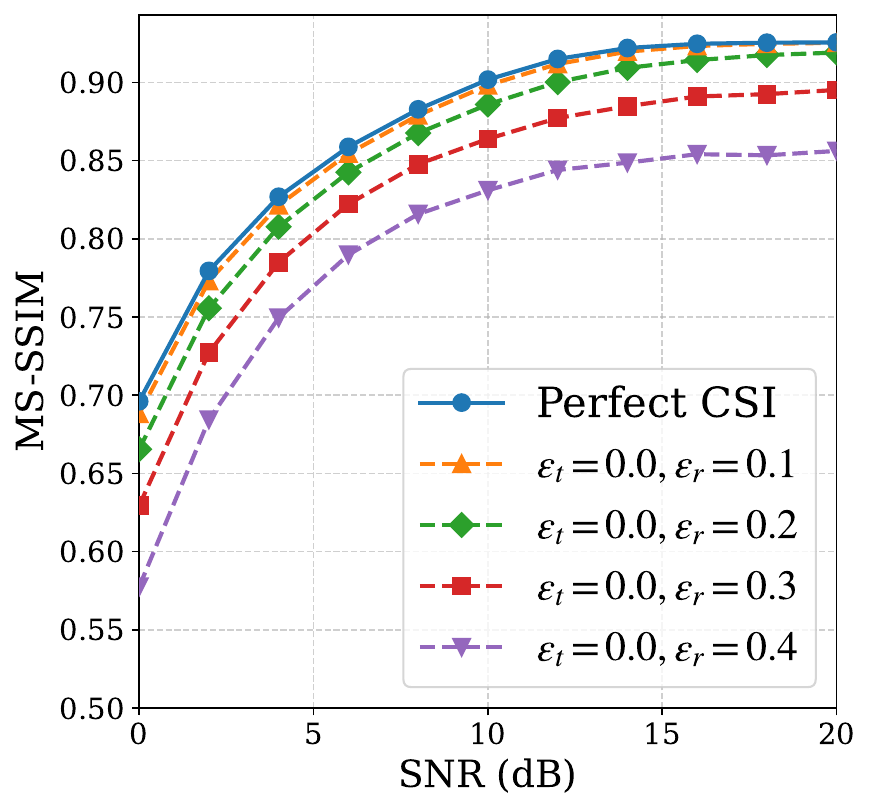}
        \subcaption{MS-SSIM under RX-side errors ($\varepsilon_t = 0$)}
        \label{fig-csi_rx_MS_SSIM}
    \end{subfigure}
    \caption{PSNR and MS-SSIM performance of the proposed SSC system under various channel estimation errors.}
    \label{fig-csi}
\end{figure}
\subsection[Comparison with More SOTA Baselines]{Comparison with More SOTA Baselines} \label{Sec-Simulation-baselines}
To further demonstrate the competitive advantages of the proposed SSC system against stronger digital semantic communication benchmarks, we implement and evaluate two additional state-of-the-art (SOTA) digital baselines, namely MOC-RVQ \cite{ref-MOC-RVQ} and ESC-MVQ \cite{ref-R3-ESC-MVQ}. 
To ensure a fair comparison under an identical compression rate of 2.0 bpp, both baselines utilize the same SwinJSCC backbone and FA-IM modulation. 
Specifically, for the digital components, MOC-RVQ incorporates an 8-head structure with a codebook size of 16. 
Meanwhile, ESC-MVQ is trained with 4 parallel codebooks of size 65536, under the assumption of perfect CSI at the transmitter. 
The comparative PSNR and MS-SSIM results across varying SNRs are presented in \figref{fig-baseline_psnr_ms_ssim}. 
As observed, the proposed SSC system consistently and significantly outperforms both MOC-RVQ and ESC-MVQ across the entire SNR range in terms of both metrics, benefiting from the proposed semantic stream splitting scheme and the tailored training strategy. 
In the low-SNR region, SSC exhibits a massive performance gain over ESC-MVQ, which suffers from severe degradation at lower channel qualities. 
Meanwhile, within the medium-SNR range, the competitive edge of SSC over MOC-RVQ becomes even more distinct.
Notably, this superior reconstruction fidelity is achieved with a drastically lower storage footprint. 
Specifically, under this configuration, MOC-RVQ incurs an 8-fold increase in codebook storage overhead compared to our SSC, while ESC-MVQ requires a staggering 16384-fold larger storage footprint. 
It is worth noting that while MOC-RVQ and ESC-MVQ focus heavily on enhancing the source-side quantization representation capability, our proposed SSC framework focuses on the cross-layer co-design of standard quantization and physical-layer IM. 
This conceptual distinction highlights that our framework is highly complementary, and the multi-codebook or multi-head mechanisms from these baselines can be seamlessly integrated into our SSC system in future extensions to achieve even stronger performance.

\subsection[Effects of the Imperfect CSI Estimation]{Effects of the Imperfect CSI Estimation} \label{Sec-Simulation-csi}
The semantic transmission scheme of the proposed SSC system is implemented with FA-IM. 
This practical scenario inevitably involves channel estimation errors, where CSI is required at both the transmitter for optimal sub-channel selection and the receiver for ML detection. 
Specifically, the imperfect channel matrices at the transmitter and receiver are modeled as $\mathbf{H}_{\text{err}} = \sqrt{1-\varepsilon_t^2}\mathbf{H} + \varepsilon_t \Delta \mathbf{H}$ and $\bar{\mathbf{H}}_{\text{err}} = \sqrt{1-\varepsilon_r^2}\bar{\mathbf{H}} + \varepsilon_r \Delta \bar{\mathbf{H}}$, respectively. 
Here, $\varepsilon_t$ and $\varepsilon_r$ denote the corresponding error coefficients, while $\Delta \mathbf{H}$ and $\Delta \bar{\mathbf{H}}$ represent the estimation noise following the same distribution as the true channel matrices. 
Notably, the case of $\varepsilon_t = \varepsilon_r = 0$ corresponds to the ideal scenario with perfect CSI.
\figref{fig-csi} examines the impact of these channel estimation errors on the PSNR and MS-SSIM performance of the proposed SSC system under varying channel estimation errors. 
As observed from the curves, both PSNR and MS-SSIM metrics exhibit a graceful degradation as the error coefficients increase. 
More importantly, a clear performance disparity can be identified between the transmitter and receiver imperfections. 
Specifically, $\varepsilon_t$ exerts a relatively negligible impact on the overall system performance. 
This resilience is primarily because the transmitter benefits from the substantial diversity gain provided by the abundant ports of the FA.
In sharp contrast, $\varepsilon_r$ leads to a much more pronounced degradation in both PSNR and MS-SSIM. 
Nevertheless, overall, the impact of channel estimation errors on the proposed SSC system remains within an expected and normal range, demonstrating that the system is relatively robust.

\begin{figure}[t]
    \centering
    \begin{subfigure}{0.49\columnwidth}
        \centering
        \includegraphics[width=\linewidth]{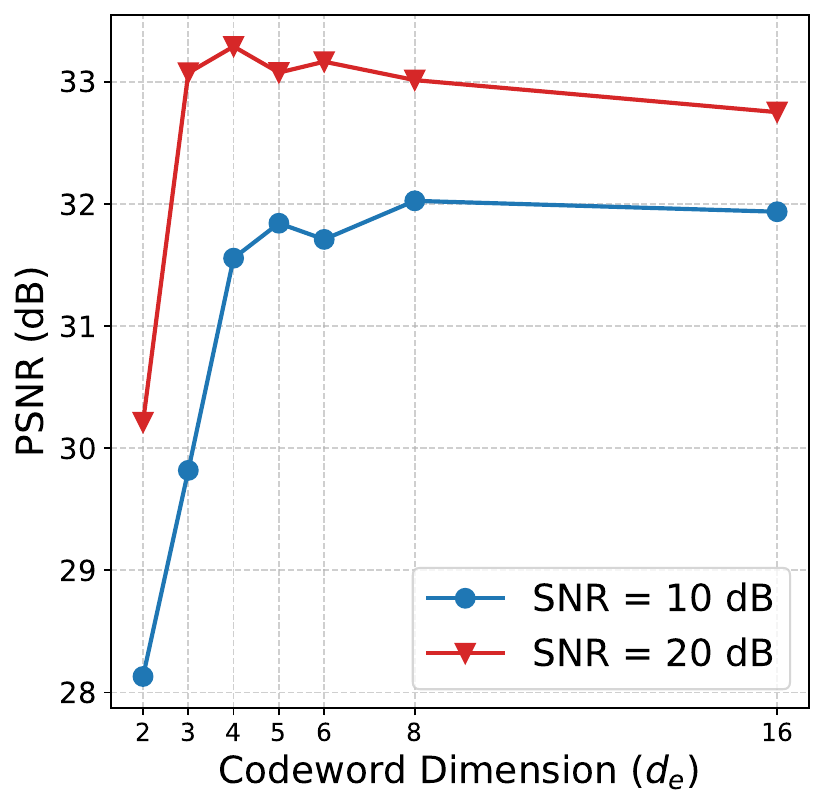}
        \subcaption{PSNR with $N_e=16$}
        \label{fig-de_16E_PSNR}
    \end{subfigure}
    \hfill 
    \begin{subfigure}{0.49\columnwidth}
        \centering
        \includegraphics[width=\linewidth]{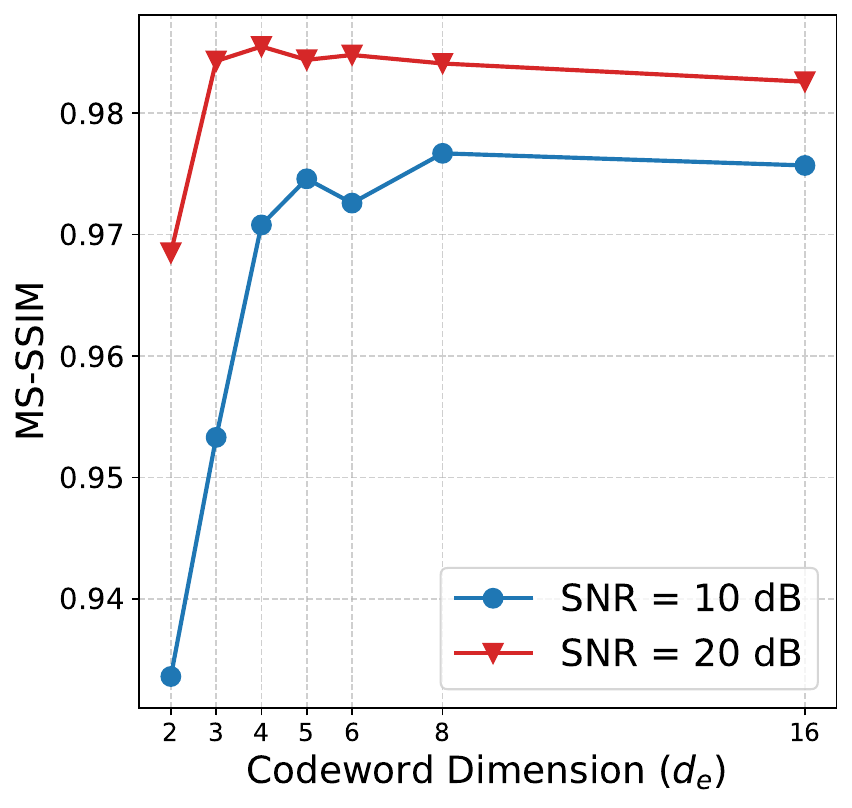}
        \subcaption{MS-SSIM with $N_e=16$}
        \label{fig-de_16E_MS_SSIM}
    \end{subfigure}
    \\
    \begin{subfigure}{0.49\columnwidth}
        \centering
        \includegraphics[width=\linewidth]{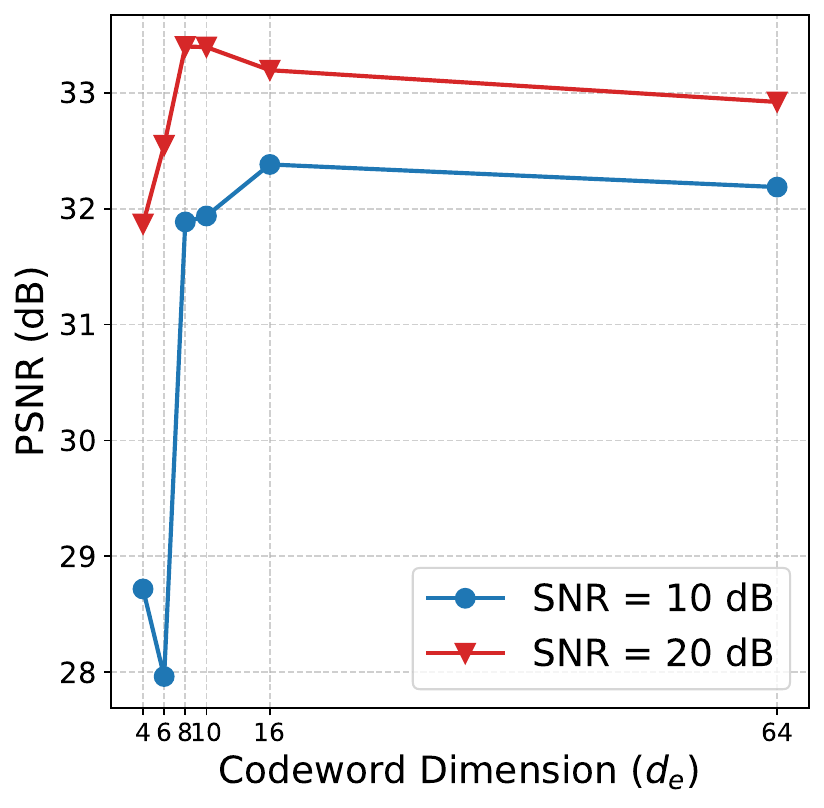}
        \subcaption{PSNR with $N_e=64$}
        \label{fig-de_64E_PSNR}
    \end{subfigure}
    \hfill 
    \begin{subfigure}{0.49\columnwidth}
        \centering
        \includegraphics[width=\linewidth]{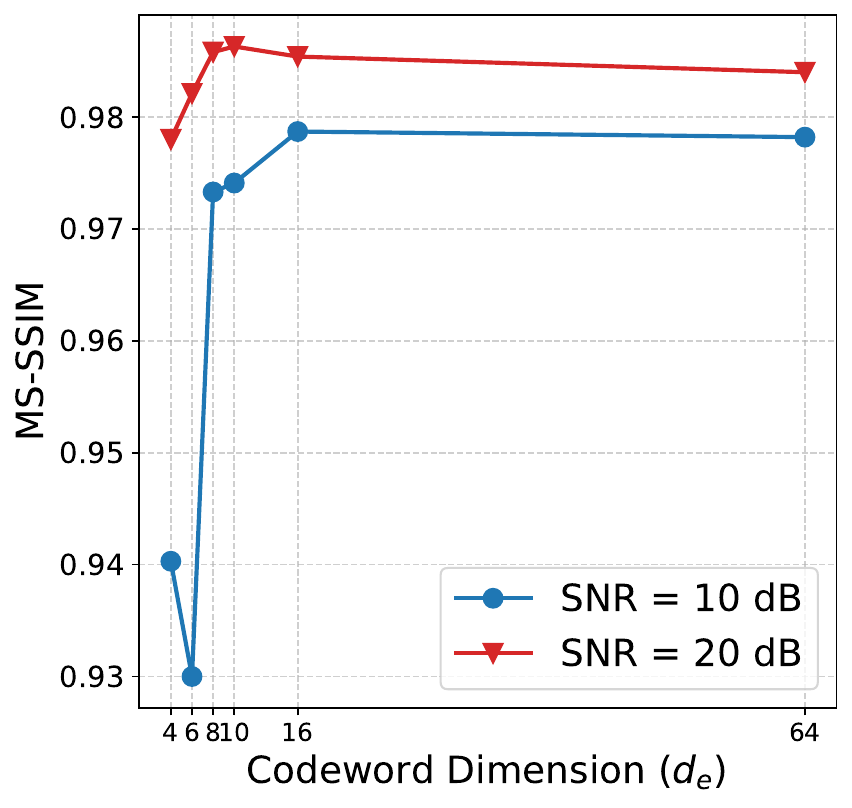}
        \subcaption{MS-SSIM with $N_e=64$}
        \label{fig-de_64E_MS_SSIM}
    \end{subfigure}
    \caption{PSNR and MS-SSIM performance of the proposed SSC system versus the codeword dimension $d_e$ under different SNRs.}
    \label{fig-de}
\end{figure}
\begin{figure}[t]
    \centering
    \begin{subfigure}{0.49\columnwidth}
        \centering
        \includegraphics[width=\linewidth]{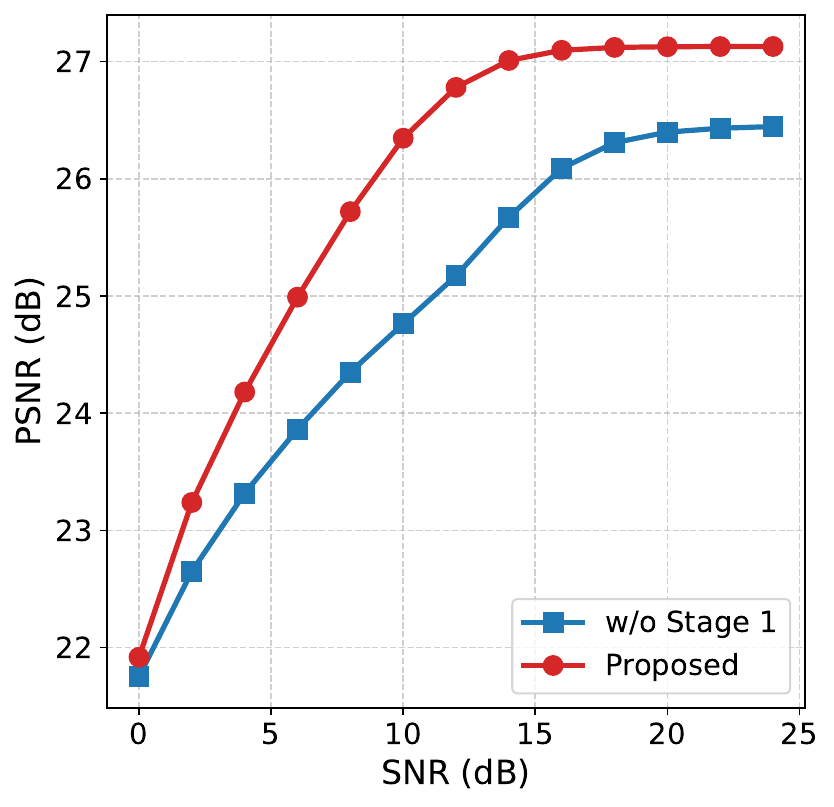}
        \subcaption{PSNR at 2.0 bpp}
        \label{fig-train_32C_PSNR}
    \end{subfigure}
    \hfill 
    \begin{subfigure}{0.49\columnwidth}
        \centering
        \includegraphics[width=\linewidth]{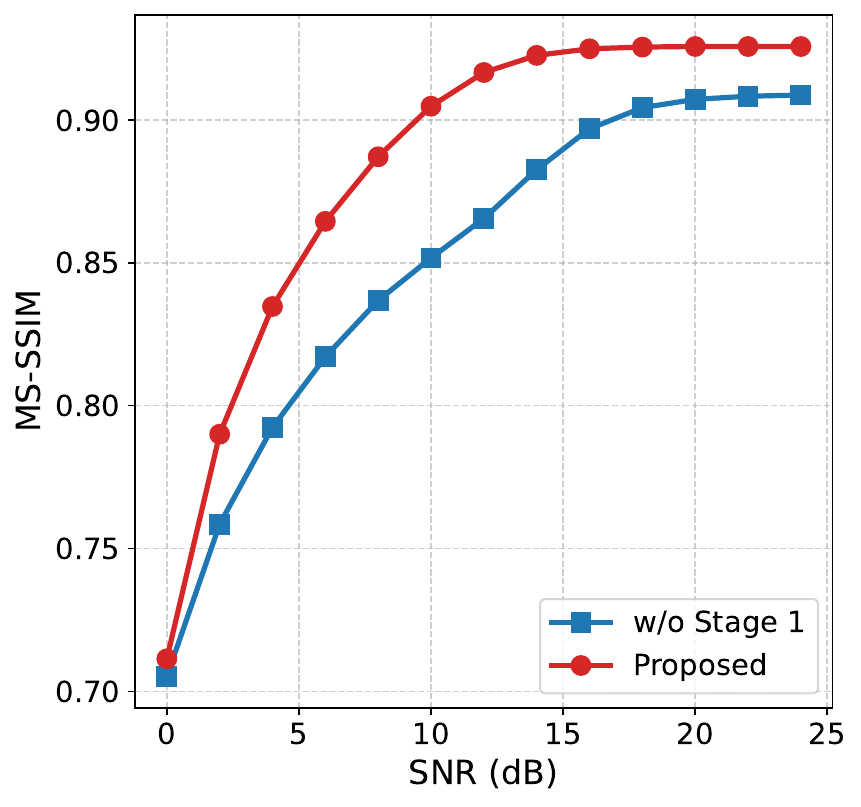}
        \subcaption{MS-SSIM at 2.0 bpp}
        \label{fig-train_32C_MS_SSIM}
    \end{subfigure}
    \\
    \begin{subfigure}{0.49\columnwidth}
        \centering
        \includegraphics[width=\linewidth]{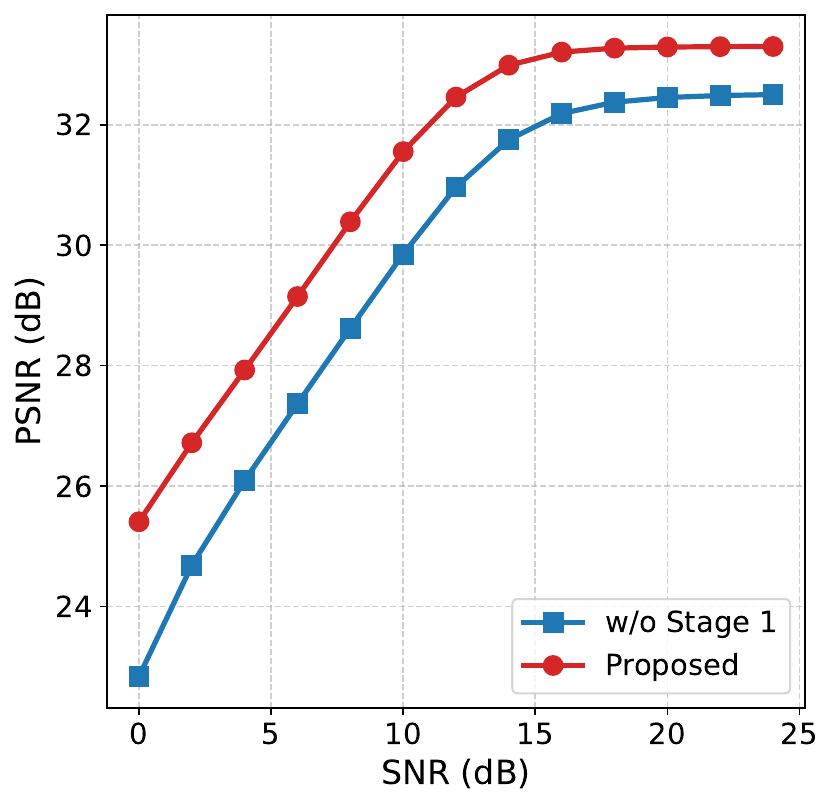}
        \subcaption{PSNR at 6.0 bpp}
        \label{fig-train_96C_PSNR}
    \end{subfigure}
    \hfill 
    \begin{subfigure}{0.49\columnwidth}
        \centering
        \includegraphics[width=\linewidth]{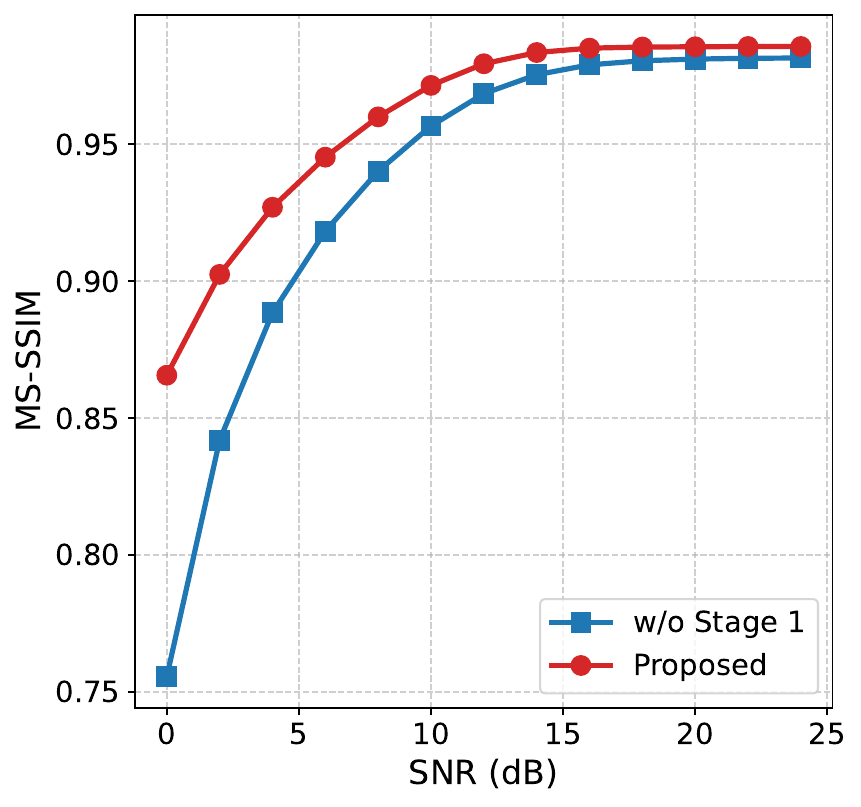}
        \subcaption{MS-SSIM at 6.0 bpp}
        \label{fig-train_96C_MS_SSIM}
    \end{subfigure}
    \caption{PSNR and MS-SSIM performance of the SSC system trained with different strategies.}
    \label{fig-train}
\end{figure}
\subsection{Evaluation of Codeword Dimension} \label{Sec-Simulation-de}
\figref{fig-de} presents the PSNR and MS-SSIM performance of the proposed SSC system with respect to the codeword dimension $d_e$ under different SNRs. 
In this experiment, the channel dimension $C_z$ and the number of RQ quantization steps $N_q$ are fixed at 96 and 4, respectively, to maintain a constant bpp.
Then, two codebook sizes are investigated, i.e., $N_e = 16$ and $N_e = 64$.
It can be observed that in the low-dimensional region (e.g., $d_e < 4$ for $N_e=16$ and $d_e < 8$ for $N_e=64$), increasing $d_e$ leads to a significant improvement in both PSNR and MS-SSIM. 
This is attributed to the enhanced representation capability of the codewords. 
A higher-dimensional space allows the codewords to capture more complex and fine-grained features from the semantic latents, thereby reducing the quantization distortion.
However, contrary to the intuition that higher dimensionality always yields better performance, the curves exhibit a distinct saturation or even a slight decline as $d_e$ continues to increase. 
For instance, with $N_e = 16$ at $\text{SNR}=10$ dB, the performance tends to saturate when $d_e$ reaches 4, whereas at $\text{SNR}=20$ dB, it peaks around $d_e = 4$ and subsequently drops as $d_e$ increases to 16. 
This phenomenon can be explained by the curse of dimensionality and the increasing difficulty in optimization during the training process.
As the dimension $d_e$ expands, learning a compact and representative codebook in such a high-dimensional space becomes challenging for the neural network.
Based on these observations, $d_e$ should be carefully tuned to balance representation capability and trainability. 
Furthermore, considering that increasing $d_e$ also imposes a higher computational burden, it is recommended to select $d_e = 4$ for $N_e = 16$ and $d_e = 8$ for $N_e = 64$ to fully unleash the potential of the proposed SSC system.

\begin{figure}[t]
    \centering
    \begin{subfigure}{0.240\textwidth}
        \centering
        \includegraphics[width=\linewidth]{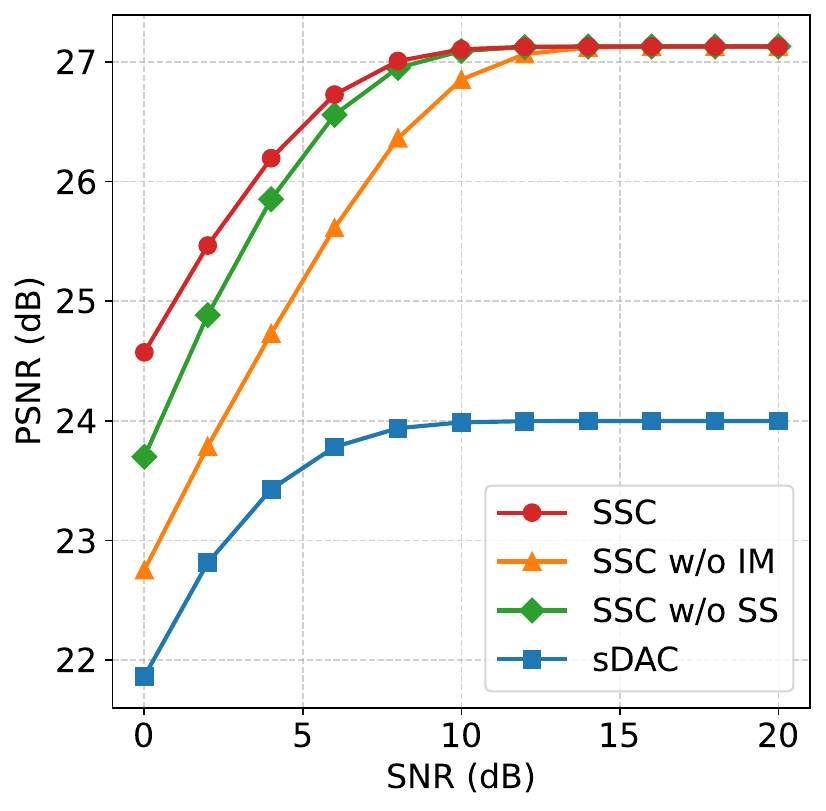}
        \subcaption{$N_s=2, M=16$}
        \label{fig-KM_psnr_1}
    \end{subfigure}
    \hfill 
    \begin{subfigure}{0.240\textwidth}
        \centering
        \includegraphics[width=\linewidth]{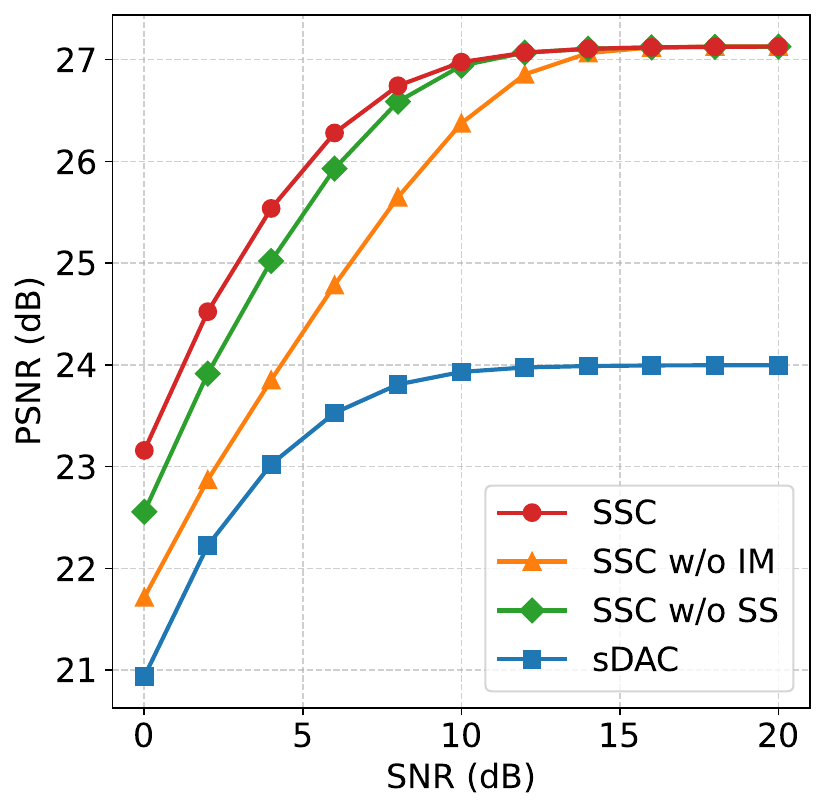}
        \subcaption{$N_s=4, M=16$}
        \label{fig-KM_psnr_2}
    \end{subfigure}
    \\ 
    \begin{subfigure}{0.240\textwidth}
        \centering
        \includegraphics[width=\linewidth]{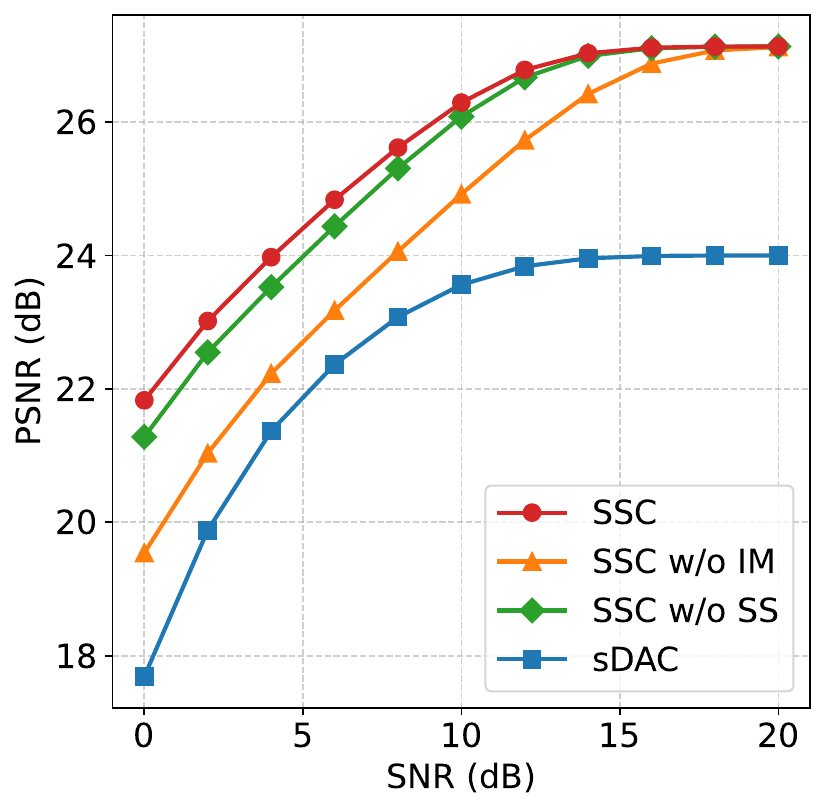}
        \subcaption{$N_s=2, M=64$}
        \label{fig-KM_psnr_3}
    \end{subfigure}\hfill
    \begin{subfigure}{0.240\textwidth}
        \centering
        \includegraphics[width=\linewidth]{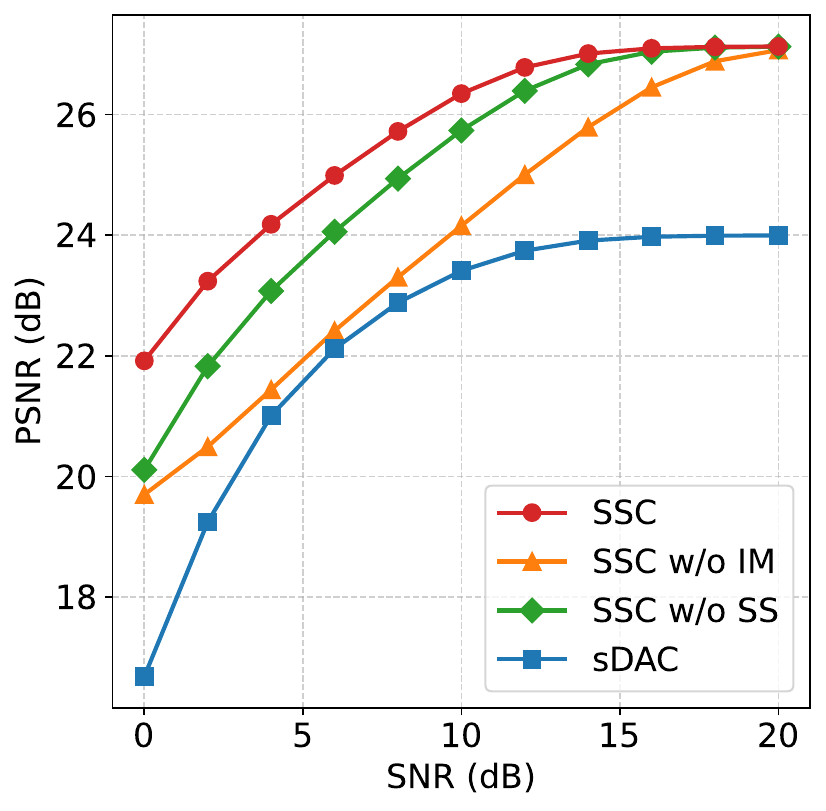}
        \subcaption{$N_s=4, M=64$}
        \label{fig-KM_psnr_4}
    \end{subfigure}
    \\ 
    \begin{subfigure}{0.240\textwidth}
        \centering
        \includegraphics[width=\linewidth]{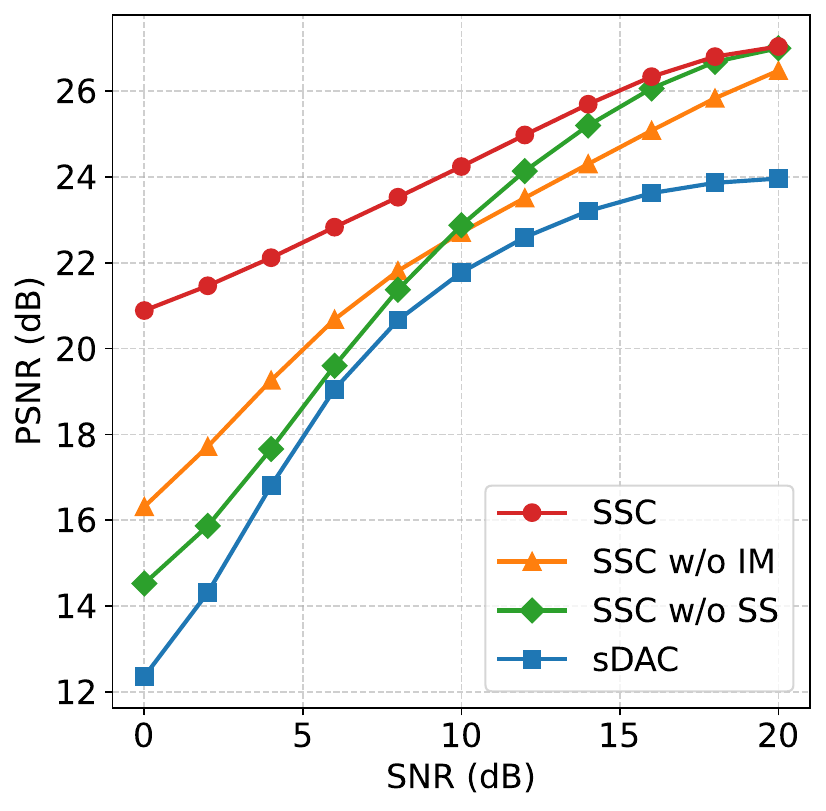}
        \subcaption{$N_s=2, M=256$}
        \label{fig-KM_psnr_5}
    \end{subfigure}
    \hfill 
    \begin{subfigure}{0.240\textwidth}
        \centering
        \includegraphics[width=\linewidth]{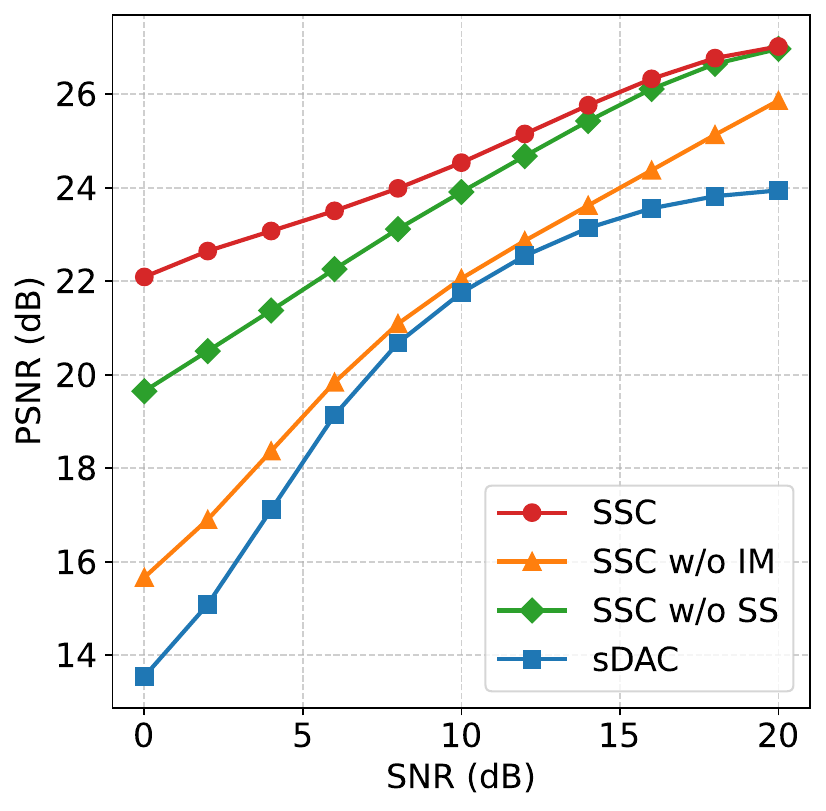}
        \subcaption{$N_s=4, M=256$}
        \label{fig-KM_psnr_6}
    \end{subfigure}
    \caption{PSNR performance comparison of the proposed SSC system and benchmarks under different FA-IM configurations: (a) $N_s=2, M=16$; (b) $N_s=4, M=16$; (c) $N_s=2, M=64$; (d) $N_s=4, M=64$; (e) $N_s=2, M=256$; (f) $N_s=4, M=256$.}
    \label{fig-KM_psnr}
\end{figure}
\begin{figure}[t]
    \centering
    \begin{subfigure}{0.240\textwidth}
        \centering
        \includegraphics[width=\linewidth]{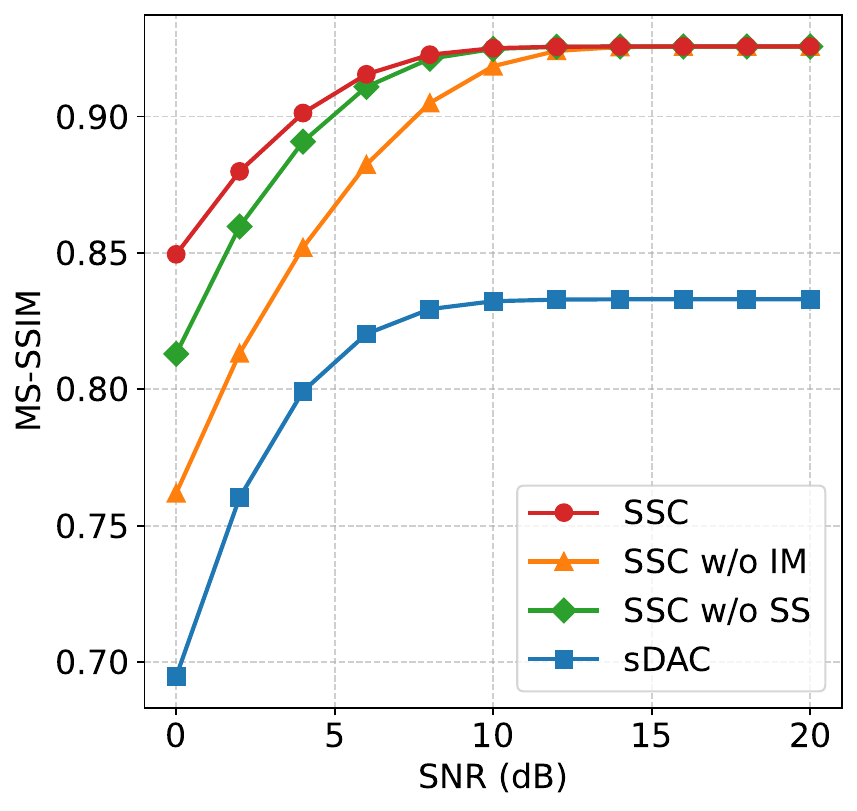}
        \subcaption{$N_s=2, M=16$}
        \label{fig-KM_ms_ssim_1}
    \end{subfigure}
    \hfill 
    \begin{subfigure}{0.240\textwidth}
        \centering
        \includegraphics[width=\linewidth]{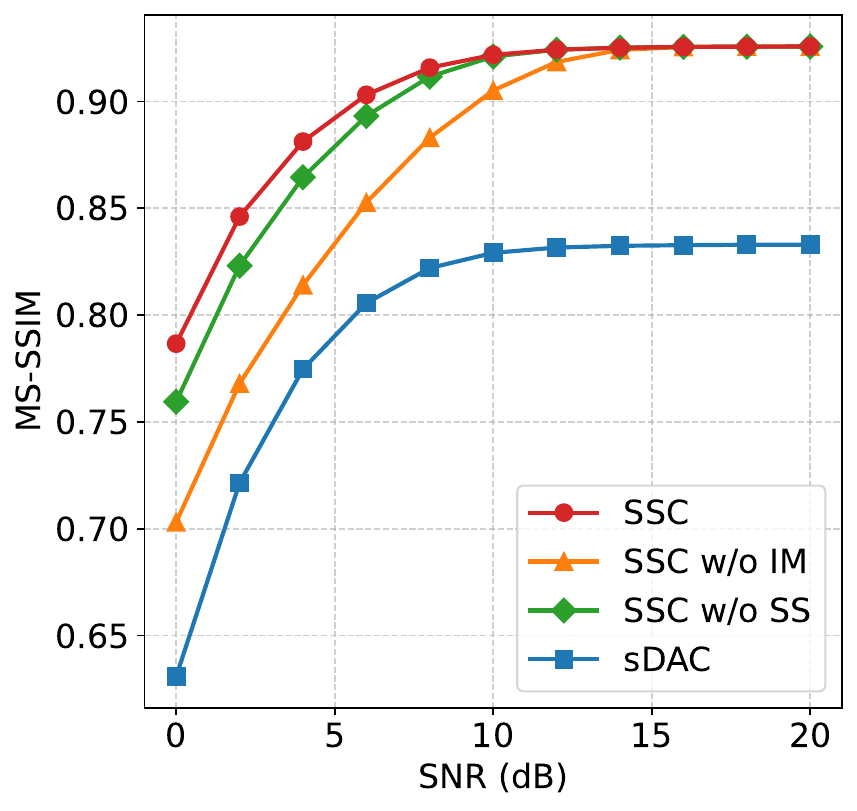}
        \subcaption{$N_s=4, M=16$}
        \label{fig-KM_ms_ssim_2}
    \end{subfigure}
    \\ 
    \begin{subfigure}{0.240\textwidth}
        \centering
        \includegraphics[width=\linewidth]{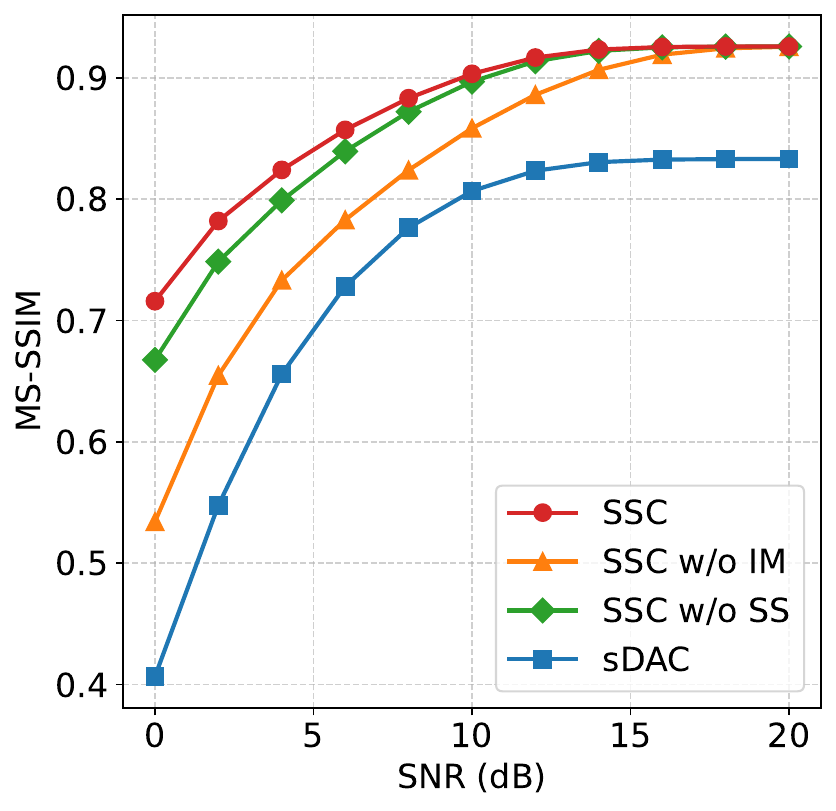}
        \subcaption{$N_s=2, M=64$}
        \label{fig-KM_ms_ssim_3}
    \end{subfigure}\hfill
    \begin{subfigure}{0.240\textwidth}
        \centering
        \includegraphics[width=\linewidth]{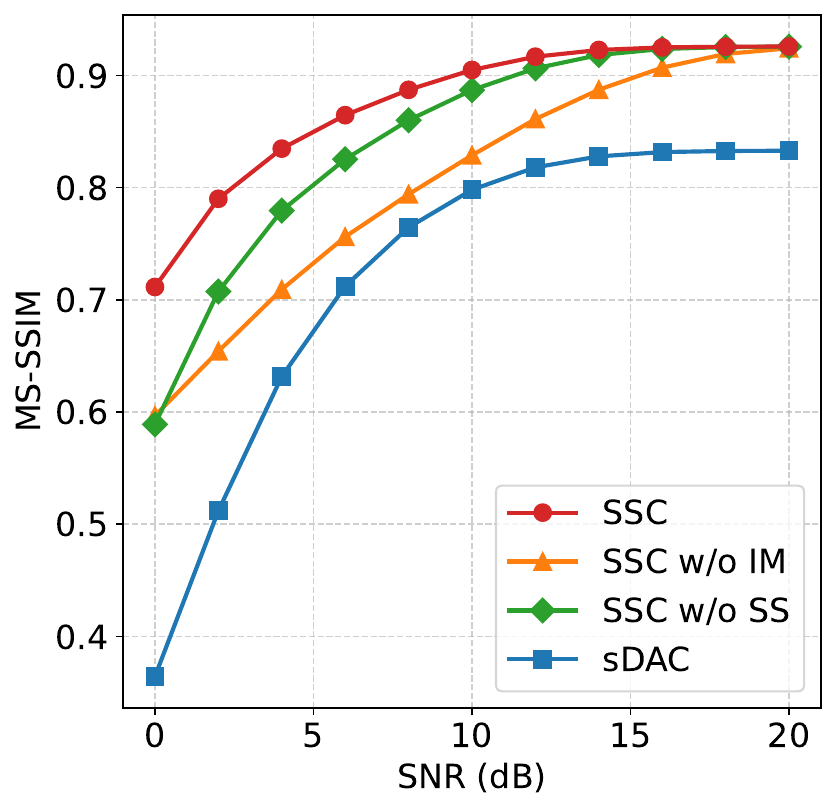}
        \subcaption{$N_s=4, M=64$}
        \label{fig-KM_ms_ssim_4}
    \end{subfigure}
    \\ 
    \begin{subfigure}{0.240\textwidth}
        \centering
        \includegraphics[width=\linewidth]{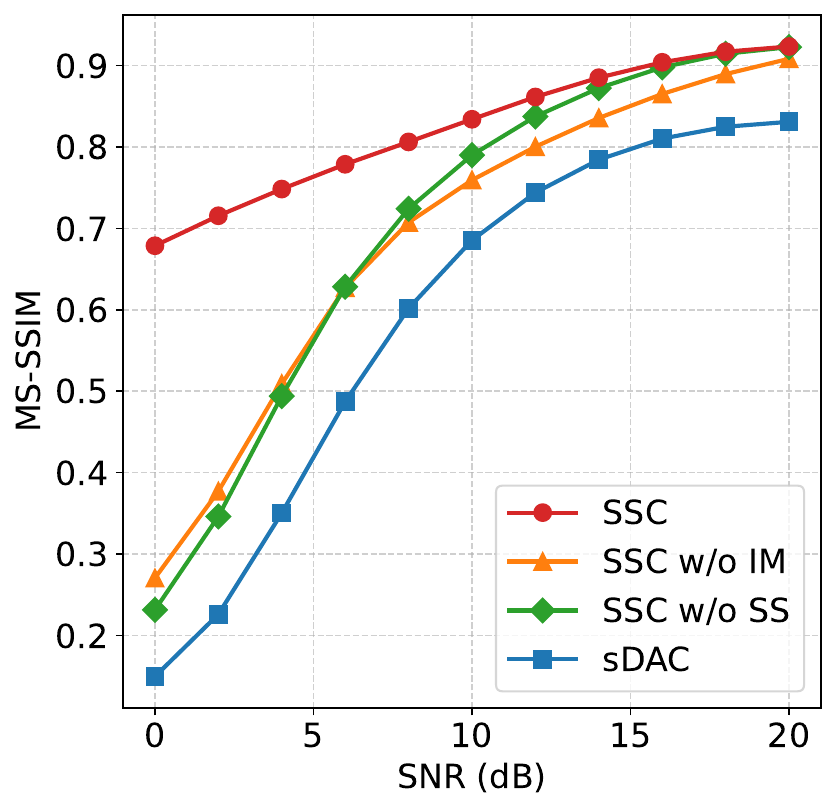}
        \subcaption{$N_s=2, M=256$}
        \label{fig-KM_ms_ssim_5}
    \end{subfigure}
    \hfill 
    \begin{subfigure}{0.240\textwidth}
        \centering
        \includegraphics[width=\linewidth]{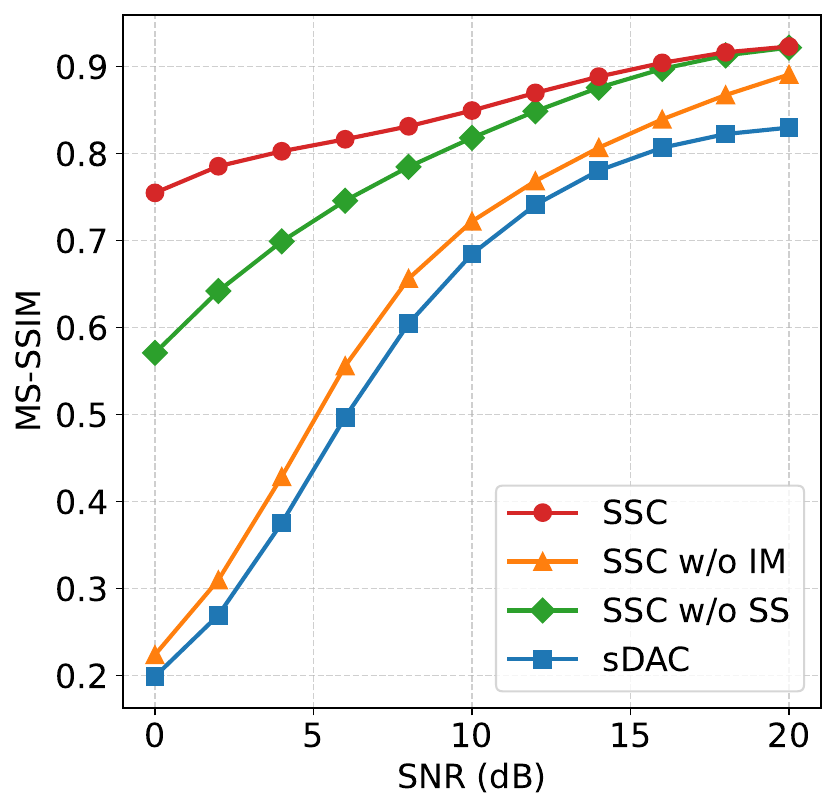}
        \subcaption{$N_s=4, M=256$}
        \label{fig-KM_ms_ssim_6}
    \end{subfigure}
    \caption{MS-SSIM performance comparison of the proposed SSC system and benchmarks under different FA-IM configurations: (a) $N_s=2, M=16$; (b) $N_s=4, M=16$; (c) $N_s=2, M=64$; (d) $N_s=4, M=64$; (e) $N_s=2, M=256$; (f) $N_s=4, M=256$.}
    \label{fig-KM_ms_ssim}
\end{figure}
\subsection{Effects of the Proposed Training Strategy} \label{Sec-Simulation-train}
Next, we evaluate the effectiveness of the proposed training strategy. 
As illustrated in \figref{fig-SSC_training}, given that Stage 0 allows for the initialization of the backbone using a pre-trained analog-domain JSCC encoder and decoder, our analysis primarily focuses on the impact of Stage 1, which involves freezing the JSCC backbone. 
\figref{fig-train} compares the performance of the SSC model with and without Stage 1, where the curves labeled `w/o Stage 1' represent a training process that skips Stage 1 and proceeds directly to global fine-tuning. 
To ensure a fair comparison, this baseline training is conducted for the same duration of $2 \times 10^5$ iterations.
We evaluated the training effectiveness of the SSC model at 2.0 bpp and 6.0 bpp, which correspond to the channel dimensions of $C_z=32$ and $C_z=96$, respectively.
As shown in \figref{fig-train}, the proposed multi-stage training strategy consistently outperforms the direct fine-tuning baseline (without Stage 1) across the entire SNR range and under different compression rates.
For instance, in the moderate SNR regime (e.g., around 10 dB), the proposed strategy achieves PSNR improvements of approximately 1.6 dB compared to the baseline at 2.0 bpp. 
In the high SNR regimes, the proposed strategy converges to a higher upper bound, indicating superior intrinsic representation capability of the model.
The significant performance degradation observed in the `w/o Stage 1' case highlights the critical role of the frozen-backbone training phase.
When skipping Stage 1, the randomly initialized RQ module and the pre-trained JSCC backbone are updated simultaneously. 
The large, erratic gradients from the untrained quantizer backpropagate to the encoder, destabilizing the well-learned analog semantic feature space. 
This phenomenon leads to catastrophic forgetting of the representations learned by the pre-trained JSCC backbone.
In contrast, by freezing the backbone in Stage 1, the proposed strategy forces the RQ module to adapt itself to the existing semantic manifold. 
This creates a stable initial alignment between the continuous latent space and the discrete codebook. 
Consequently, the subsequent global fine-tuning (Stage 2) can start from a robust operating point, leading to faster convergence and superior final reconstruction quality.
These results confirm that the proposed three-stage training strategy is a necessary component to fully leverage the potential of the JSCC backbone and the RQ module within the SSC framework.

\subsection{Evaluation of FA-IM Parameters} \label{Sec-Simulation-IM}
\figref{fig-KM_psnr} and \figref{fig-KM_ms_ssim} illustrate the comprehensive performance comparison between the proposed SSC system and three benchmarks  under different FA-IM configurations. 
These evaluations are conducted by varying the number of ports employed for IM, $N_s \in \{2, 4\}$, and the constellation size $M \in \{16, 64, 256\}$.
As can be observed, the proposed SSC system consistently exhibits superior performance compared to all benchmarks across all tested configurations.
Specifically, as $M$ increases, the benchmarks suffers from noticeable degradation in the low SNR regime.
However, the proposed SSC system maintains a robust performance lead, with the advantage becoming even more pronounced at the configuration of $N_s=4$ and $M=256$.
Meanwhile, increasing $N_s$ increases the SE by introducing more index bits without densifying the constellation.
Consequently, it generally exhibits improved performance, demonstrating the efficacy of exploiting spatial domain indices to convey additional semantic information robustly.
The performance gap between the proposed SSC and the ``SSC w/o IM'' benchmark also underscores the benefits of IM. 
By offloading a portion of the semantic information to the port indices, the proposed SSC system achieves a much faster performance climb. 
Furthermore, the superiority of the proposed SSC over the ``SSC w/o SS'' baseline highlights the necessity of the proposed stream splitting strategy. 
Our semantic-aware splitting explicitly maps the most significant semantic information to the robust data stream. 
This unequal error protection mechanism ensures that the structural integrity of the image is preserved even when the constellation symbols are corrupted, yielding consistently higher PSNR and MS-SSIM scores.
Compared to the sDAC scheme, the proposed SSC system exhibits a overwhelming advantage. 
The sDAC scheme, even with FA-IM, cannot efficiently represent complex high-dimensional semantic features, resulting in early performance saturation. 
The proposed SSC leverages RQ to decompose features into a multi-stage discrete representation, enabling high-fidelity reconstruction with a compact codebook. 

\section{Conclusion} \label{Sec-Conclusion}
This paper pioneered the integration of semantic communications with FA-IM, proposing a novel SSC system. 
By leveraging RQ, the SSC system discretizes continuous semantic features into a compact hierarchical representation, effectively bridging the gap between the analog JSCC backbone and the digital transmission infrastructure. 
Furthermore, fully capitalizing on the unique characteristics of RQ and FA-IM, we designed a semantic-aware stream splitting mechanism that prioritizes the allocation of critical semantic information to the more robust FA-IM transmission dimension, thereby achieving superior semantic fidelity.
Extensive simulation results have validated the efficacy of the proposed training strategy and demonstrated that the SSC system consistently outperforms benchmarks in terms of PSNR and MS-SSIM.
Collectively, these findings underscore the potential of the proposed SSC system as a spectrum-efficient solution for future 6G intelligent communications.



\bibliographystyle{IEEEtran}
\bibliography{IEEEabrv,mybib}


 





\end{document}